\documentclass[11pt]{article}
\usepackage{epsfig} 
\newcommand{\sect}[1]{ \section{#1} \setcounter{equation}{0} }

\newcommand{\pslash}{p \! \! \! /}

\newcommand{\half}{\mbox{\small{$\frac{1}{2}$}}} 
\newcommand{\Nc}{N_{\!c}} 
\newcommand{\Nf}{N_{\!f}} 
\newcommand{\MSbar}{\overline{\mbox{MS}}}

\begin{document}
\title{Three loop anomalous dimensions of higher moments of the non-singlet
twist-$2$ Wilson and transversity operators in the $\MSbar$ and RI${}^\prime$ 
schemes} 
\author{J.A. Gracey, \\ Theoretical Physics Division, \\ 
Department of Mathematical Sciences, \\ University of Liverpool, \\ P.O. Box 
147, \\ Liverpool, \\ L69 3BX, \\ United Kingdom.} 
\date{} 
\maketitle 
\vspace{5cm} 
\noindent 
{\bf Abstract.} We compute the anomalous dimension of the third and fourth 
moments of the flavour non-singlet twist-$2$ Wilson and transversity operators 
at three loops in both the $\MSbar$ and RI$^\prime$ schemes. To assist with the
extraction of estimates of matrix elements computed using lattice 
regularization, the finite parts of the Green's function where the operator is
inserted in a quark $2$-point function are also provided at three loops in both
schemes. 

\vspace{-17cm}
\hspace{13.5cm}
{\bf LTH 718} 

\newpage

\sect{Introduction.} 
In calculations relating to deep inelastic scattering the operator product 
expansion plays an important role in allowing one to evaluate the underlying
current correlators. In essence there are two parts to the formalism. The first
is the basis of gauge invariant operators into which the current correlators
are decomposed and the other is the process dependent Wilson coefficients. For 
high energy experiments the dominant set of operators are those of leading 
twist defined as the difference between dimension and spin. The Wilson 
coefficients are determined from the specific process of interest and are 
computed in perturbation theory. To understand the physics at various energies 
one requires the solution of the underlying renormalization group equation at 
some particular loop approximation and necessary for this are the anomalous 
dimensions of the operators of the basis. With the increased precision now 
demanded by experiments the goal in recent years has been to ascertain the 
anomalous dimensions at three loops as an analytic function of the moment, $n$,
of the operators. This has been achieved by the magnificent computation of
\cite{1,2,3,4} for the twist-$2$ flavour non-singlet and singlet Wilson 
operators in the $\MSbar$ scheme as well as the Wilson coefficients to the same
order which extended the lower loop results of \cite{5,6,7}. Hence the 
{\em full} two loop renormalization group evolution has been determined. 
However, one feature which cannot be deduced from perturbative techniques is 
the underlying matrix element which is non-perturbative in nature and has to be
deduced by use of the lattice where various moments of the matrix element
have been determined for low moments. See, for example, the ongoing work of the
QCDSF collaboration, \cite{8,9,10,11,12,13}, and others, \cite{14,15,16,17}. 
However, in making measurements of matrix elements, which ultimately will 
provide accurate predictions for experiments, there are several issues. 

First, to reduce computation time and consequently cost, the matrix elements
are determined on the lattice in a renormalization scheme geared for this
problem which is known as the regularization invariant (RI) scheme and its
variation called the modified regularization invariant (RI$^\prime$) scheme,
\cite{18}. Both are mass dependent renormalization schemes. However, when the
results are extracted in this scheme they need to be converted to the reference
$\MSbar$ scheme which is a mass independent scheme. An early example of the 
application of this approach is given in \cite{19,20}. Second, to ensure that 
the results are credible when evolved to large energy they must match the 
perturbative result for the matrix element in the {\em same} renormalization 
scheme. There are several ways of achieving this. One is to use a 
non-perturbative approach such as the Schr\"{o}dinger functional method, 
\cite{21}. The other is to compute the matrix element to as many orders as 
possible in conventional perturbation theory and then match the lattice results
to the explicit perturbative results in the same renormalization scheme. 
Previously that has been the approach of Chetyrkin and R\'{e}tey, \cite{20}, 
and \cite{22,23}. Specifically, various quark currents have been considered 
including the tensor current as well as the second moment of the flavour 
non-singlet Wilson and transversity operators. The latter was originally 
introduced in \cite{24,25,26} and relates to the probability of finding a quark
in a transversely polarized nucleon polarized parallel to the nucleon versus 
that of the nucleon in the antiparallel polarization. The results of 
\cite{20,22,23} have been important in the matrix element lattice computations 
of \cite{11,12,13,14,15,16,17}. Necessary for the three loop perturbative 
calculations of the Green's functions with the operator inserted has been the 
full three loop renormalization of QCD in the RI$^\prime$ scheme in an 
arbitrary linear covariant gauge, \cite{20,22}. Given that there has been an 
advance in computing technology in recent years, it transpires that it is now 
feasible to measure higher moments of the underlying matrix elements on the
lattice since essentially there has been a significant improvement in 
numerically isolating a clear signal. Therefore to assist the full matching 
procedure in the ultraviolet region to produce {\em accurate} estimates, it is 
necessary to extend the approach of \cite{22,23} to higher moments of these two
classes of operators. This is the purpose of this article where we will 
consider the third and fourth moments of the flavour non-singlet Wilson and 
transversity operators in an arbitrary linear covariant gauge. There will be 
two parts to this venture. The first is the determination of the anomalous 
dimensions of the operators at three loops in $\MSbar$ and RI$^\prime$. Whilst 
the second is to produce to the same loop order the value of the Green's 
function involving the operator itself inserted in a quark $2$-point function 
in both schemes. Although it is worth noting that for the lattice, only the 
Landau gauge results are relevant since that is the gauge in which lattice 
measurements are made. The full arbitrary gauge calculation, being more 
complete, is actually important for internal checks on the loop computations. 
Whilst the even moments are appropriate for deep inelastic scattering 
experiments, the odd moments are accessible on the lattice and can serve the 
purpose of a testing ground for technical issues for higher moment lattice 
matching. Whilst the Wilson operator anomalous dimensions are known already at 
three loops in $\MSbar$, \cite{1,2}, we note that what is required are the 
values of the specific Green's function with the operator inserted which has 
not been determined previously. In also considering the transversity operator, 
we will deduce {\em new} anomalous dimensions at three loops in both $\MSbar$ 
and RI$^\prime$ beyond the earlier two loop calculation of 
\cite{27,28,29,30,31}. Moreover, since we will be using symbolic manipulation 
and computer algebra tools and given that the Wilson and transversity operators
are both formally similar, the actual computations are efficiently performed 
through the same computer programmes.

The paper is organised as follows. In section $2$ we review the basics of the
RI$^\prime$ scheme and introduce the properties of the operators we will
consider. The full three loop anomalous dimensions for these operators are
given in both schemes in section $3$, whilst the same information for the
underlying Green's functions are provided in section $4$ including the
restriction to the colour group $SU(3)$ and the Landau gauge. Section $5$
records the explicit functions which convert all the three loop anomalous 
dimensions between the $\MSbar$ and RI$^\prime$ schemes with concluding remarks
given in section $6$. Several appendices summarize the projection of the 
Green's function onto the basis of independent Lorentz tensors sharing the same
symmetry properties as the corresponding operator, as well as the construction
of the full operator with these symmetries.

\sect{RI$^\prime$ scheme.} 
In this section we discuss the definition of the RI$^\prime$ scheme and 
properties of the operators we are interested in. First, we recall that the
standard renormalization scheme is the MS scheme, \cite{32}, where the poles 
with respect to the regulator are absorbed into the renormalization constants. 
Its widely used extension, $\MSbar$, which is also a mass independent scheme
additionally absorbs the finite part $\ln ( 4\pi e^{-\gamma} )$ where $\gamma$ 
is the Euler-Mascheroni constant, \cite{33}. By contrast the regularization 
invariant schemes, \cite{18}, are mass dependent schemes which from the point 
of view of the Lagrangian is the scheme where the quark $2$-point function not 
only is rendered finite but also chosen to be unity through the RI definition  
\begin{equation} 
\left. \lim_{\epsilon \rightarrow 0} \left[ \frac{1}{4d} \mbox{tr} \left( 
Z^{\mbox{\footnotesize{RI}}}_\psi \gamma^\mu \frac{\partial ~}{\partial p^\mu} 
\Sigma_\psi(p) \right) \right] \right|_{p^2 \, = \, \mu^2} ~=~ 1 ~. 
\label{ridef} 
\end{equation}  
where $\Sigma_\psi(p)$ is the quark $2$-point function with external momentum 
$p$, $Z^{\mbox{\footnotesize{${\cal R}$}}}_\psi$ is the quark wave function 
renormalization constant in the ${\cal R}$ renormalization scheme and $\mu$ is
the scale introduced to ensure that the coupling constant remains dimensionless
in $d$ dimensions when using dimensional regularization with
$d$~$=$~$4$~$-$~$2\epsilon$. However, in practice since taking a derivative is 
(financially) costly on the lattice, one takes a variation of this definition 
(\ref{ridef}), to define the RI$^\prime$ scheme which does not involve 
differentiation, \cite{18}, which is 
\begin{equation} 
\left. \lim_{\epsilon \rightarrow 0} \left[ 
Z^{\mbox{\footnotesize{RI$^\prime$}}}_\psi \Sigma_\psi(p) \right] 
\right|_{p^2 \, = \, \mu^2} ~=~ \pslash ~. 
\label{ripdef} 
\end{equation}  
Although this is primarily the key to defining the scheme on the lattice as
well as the continuum, the full three loop renormalization of QCD has been
performed in an arbitrary linear covariant gauge and colour group in \cite{22}.
Additionally part of the four loop renormalization has been performed for
$SU(\Nc)$ in \cite{20}. However, in \cite{22} the other field $2$-point 
functions were defined in an analogous way to (\ref{ripdef}). Further, by 
contrast the $3$-point functions were renormalized according to the usual 
$\MSbar$ condition. So that those Green's functions were not constrained to be 
unity. Consequently, the relationship between the variables in both schemes 
were established to three loops and specifically are, \cite{22},  
\begin{equation}   
a_{\mbox{\footnotesize{RI$^\prime$}}} ~=~ 
a_{\mbox{\footnotesize{$\MSbar$}}} ~+~ O \left( 
a_{\mbox{\footnotesize{$\MSbar$}}}^5 \right) 
\label{cccon} 
\end{equation}  
where $a$~$=$~$g^2/(16\pi^2)$ in terms of the coupling constant $g$ in the
definition of the covariant derivative $D_\mu$, and for the arbitrary linear 
covariant gauge parameter $\alpha$,  
\begin{eqnarray}
\alpha_{\mbox{\footnotesize{RI$^\prime$}}} 
&=& \left[ 1 + \left( \left( - 9 \alpha_{\mbox{\footnotesize{$\MSbar$}}}^2 
- 18 \alpha_{\mbox{\footnotesize{$\MSbar$}}} - 97 \right) C_A + 80 T_F \Nf 
\right) \frac{a_{\mbox{\footnotesize{$\MSbar$}}}}{36} \right. \nonumber \\ 
&& \left. ~+~ \left( \left( 18 \alpha_{\mbox{\footnotesize{$\MSbar$}}}^4 
- 18 \alpha_{\mbox{\footnotesize{$\MSbar$}}}^3 
+ 190 \alpha_{\mbox{\footnotesize{$\MSbar$}}}^2 
- 576 \zeta(3) \alpha_{\mbox{\footnotesize{$\MSbar$}}} 
+ 463 \alpha_{\mbox{\footnotesize{$\MSbar$}}} + 864 \zeta(3) - 7143 \right) 
C_A^2 \right. \right. \nonumber \\ 
&& \left. \left. ~~~~~~~+~ \left( -~ 320 
\alpha_{\mbox{\footnotesize{$\MSbar$}}}^2 
- 320 \alpha_{\mbox{\footnotesize{$\MSbar$}}} + 2304 \zeta(3) 
+ 4248 \right) C_A T_F \Nf \right. \right. \nonumber \\ 
&& \left. \left. ~~~~~~~+~ \frac{}{} \left( -~ 4608 \zeta(3) 
+ 5280 \right) C_F T_F \Nf \right) 
\frac{a^2_{\mbox{\footnotesize{$\MSbar$}}}}{288} \right. \nonumber \\
&& \left. ~+~ \left( \left( ~-~ 486 \alpha_{\mbox{\footnotesize{$\MSbar$}}}^6 
+ 1944 \alpha_{\mbox{\footnotesize{$\MSbar$}}}^5 
+ 4212 \zeta(3) \alpha_{\mbox{\footnotesize{$\MSbar$}}}^4 
- 5670 \zeta(5) \alpha_{\mbox{\footnotesize{$\MSbar$}}}^4 
- 18792 \alpha_{\mbox{\footnotesize{$\MSbar$}}}^4 \right. \right. \right.   
\nonumber \\
&& \left. \left. \left. ~~~~~~~~+~ 48276 \zeta(3) 
\alpha_{\mbox{\footnotesize{$\MSbar$}}}^3 
- 6480 \zeta(5) \alpha_{\mbox{\footnotesize{$\MSbar$}}}^3 
- 75951 \alpha_{\mbox{\footnotesize{$\MSbar$}}}^3 
- 52164 \zeta(3) \alpha_{\mbox{\footnotesize{$\MSbar$}}}^2 
\right. \right. \right. \nonumber \\
&& \left. \left. \left. ~~~~~~~~+~ 2916 \zeta(4) 
\alpha_{\mbox{\footnotesize{$\MSbar$}}}^2 
+ 124740 \zeta(5) \alpha_{\mbox{\footnotesize{$\MSbar$}}}^2 
+ 92505 \alpha_{\mbox{\footnotesize{$\MSbar$}}}^2
- 1303668 \zeta(3) \alpha_{\mbox{\footnotesize{$\MSbar$}}} 
\right. \right. \right. \nonumber \\
&& \left. \left. \left. ~~~~~~~~+~ 11664 \zeta(4) 
\alpha_{\mbox{\footnotesize{$\MSbar$}}} 
+ 447120 \zeta(5) \alpha_{\mbox{\footnotesize{$\MSbar$}}} 
+ 354807 \alpha_{\mbox{\footnotesize{$\MSbar$}}} 
+ 2007504 \zeta(3) 
\right. \right. \right. \nonumber \\
&& \left. \left. \left. ~~~~~~~~+~ 8748 \zeta(4) + 1138050 \zeta(5) 
- 10221367 \right) C_A^3 \right. \right. \nonumber \\
&& \left. \left. ~~~~~~~+~ \left( 
12960 \alpha_{\mbox{\footnotesize{$\MSbar$}}}^4 
- 8640 \alpha_{\mbox{\footnotesize{$\MSbar$}}}^3 
- 129600 \zeta(3) \alpha_{\mbox{\footnotesize{$\MSbar$}}}^2 
- 147288 \alpha_{\mbox{\footnotesize{$\MSbar$}}}^2
\right. \right. \right. \nonumber \\
&& \left. \left. \left. ~~~~~~~~~~~~+~ 698112 \zeta(3) 
\alpha_{\mbox{\footnotesize{$\MSbar$}}} - 312336 
\alpha_{\mbox{\footnotesize{$\MSbar$}}} + 1505088 \zeta(3) - 279936 \zeta(4)
\right. \right. \right. \nonumber \\
&& \left. \left. \left. ~~~~~~~~~~~~-~ 1658880 \zeta(5) + 9236488 \right) 
C_A^2 T_F \Nf  \right. \right. \nonumber \\
&& \left. \left. ~~~~~~~+~ \left( 248832 \zeta(3) 
\alpha_{\mbox{\footnotesize{$\MSbar$}}}^2 
- 285120 \alpha_{\mbox{\footnotesize{$\MSbar$}}}^2 
+ 248832 \zeta(3) \alpha_{\mbox{\footnotesize{$\MSbar$}}} 
- 285120 \alpha_{\mbox{\footnotesize{$\MSbar$}}} 
\right. \right. \right. \nonumber \\
&& \left. \left. \left. ~~~~~~~~~~~~-~ 5156352 \zeta(3) + 373248 \zeta(4)
- 2488320 \zeta(5) + 9293664 \right) C_A C_F T_F \Nf \right. \right. 
\nonumber \\
&& \left. \left. ~~~~~~~+~ \left( ~-~ 38400 
\alpha_{\mbox{\footnotesize{$\MSbar$}}}^2 
- 221184 \zeta(3) \alpha_{\mbox{\footnotesize{$\MSbar$}}} 
+ 55296 \alpha_{\mbox{\footnotesize{$\MSbar$}}}
\right. \right. \right. \nonumber \\
&& \left. \left. \left. ~~~~~~~~~~~~~~\,-~ 884736 \zeta(3) - 1343872 \right) 
C_A T_F^2 \Nf^2 \right. \right. \nonumber \\
&& \left. \left. ~~~~~~~+~ \left( ~-~ 3068928 \zeta(3) + 4976640 \zeta(5) 
- 988416 \right) C_F^2 T_F \Nf \right. \right. \nonumber \\
&& \left. \left. ~~~~~~~+~ \left( 2101248 \zeta(3) - 2842368 \right) 
C_F T_F^2 \Nf^2 \right) 
\frac{a^3_{\mbox{\footnotesize{$\MSbar$}}}}{31104} \right] 
\alpha_{\mbox{\footnotesize{$\MSbar$}}} ~+~ O \left( 
a^4_{\mbox{\footnotesize{$\MSbar$}}} \right) 
\label{alpcon}
\end{eqnarray}
where $T_F$, $C_F$ and $C_A$ are the usual group theory factors defined by
\begin{equation}
\mbox{Tr} \left( T^a T^b \right) ~=~ T_F \delta^{ab} ~~,~~ T^a T^a ~=~ 
C_F I ~~,~~ f^{acd} f^{bcd} ~=~ C_A \delta^{ab}
\end{equation}
for colour group generators $T^a$, $\zeta(n)$ is the Riemann zeta function and 
the scheme of the variable is indicated as a subscript. Clearly only in the 
Landau gauge, where $\alpha$~$=$~$0$, are the variables equivalent. Although 
ultimately we are interested in the Landau gauge for the lattice matching, we 
have chosen to compute in an arbitrary linear covariant gauge as the extra 
$\alpha$ dependence, evident in expressions such as (\ref{alpcon}), will 
provide a central checking tool in the loop computations. For instance, in a 
mass independent renormalization scheme the anomalous dimension of a gauge 
invariant operator is independent of the gauge parameter. Though this is not
the case for mass dependent schemes such as RI$^\prime$ which will be apparent
in our explicit results.

More specifically the non-singlet operators we will focus on are
\begin{eqnarray} 
{\cal O}_W^{\nu_1\ldots\nu_n} &=& {\cal S} \bar{\psi} \gamma^{\nu_1} D^{\nu_2} 
\ldots D^{\nu_n} \psi \nonumber \\ 
{\cal O}_T^{\mu\nu_1\ldots\nu_n} &=& {\cal S} \bar{\psi} \sigma^{\mu\nu_1} 
D^{\nu_2} \ldots D^{\nu_n} \psi 
\end{eqnarray} 
with $n$~$=$~$3$ and $4$ where $\sigma^{\mu\nu}$~$=$~$\half [ \gamma^\mu, 
\gamma^\nu ]$ and is antisymmetric in its Lorentz indices. The operator 
${\cal S}$ in both sets denotes the symmetrization of the Lorentz indices 
$\{\nu_1,\ldots,\nu_n\}$ and the selection of the traceless part according to 
slightly different criteria for both cases. For the Wilson operators this is
\begin{equation} 
\eta_{\nu_i\nu_j}{\cal O}_W^{\nu_1\ldots\nu_i\ldots\nu_j\ldots\nu_n} ~=~ 0 ~.  
\end{equation}
Whilst for the transversity operators, \cite{29}, 
\begin{eqnarray}
\eta_{\mu\nu_i}{\cal O}_T^{\mu\nu_1\ldots\nu_i\ldots\nu_n} &=& 0 ~~~~ 
(i ~\geq~ 2) \nonumber \\
\eta_{\nu_i\nu_j}{\cal O}_T^{\mu\nu_1\ldots\nu_i\ldots\nu_j\ldots\nu_n} &=& 
0 ~.  
\end{eqnarray} 
Therefore, the transversity operator has formally one less traceless condition  
compared to the Wilson operators. 

For the renormalization of an operator in a quark $2$-point function, which
will be either $\langle \psi(-p) {\cal O}_W^{\mu_1\ldots\mu_n}(0) \bar{\psi}(p)
\rangle$ or $\langle \psi(-p) {\cal O}_T^{\mu\nu_1\ldots\nu_n}(0) \bar{\psi}(p)
\rangle$, the RI$^\prime$ scheme definition is similar to (\ref{ripdef}), 
\cite{18,19,20,22,23}. However, as the operators we will consider will carry 
Lorentz indices this $2$-point function will decompose into several invariant 
amplitudes which may or may not have a tree $(T)$ or Born term. For those 
amplitudes which have a tree term, the RI$^\prime$ scheme definition is, 
\cite{23},  
\begin{equation}  
\left. \lim_{\epsilon \, \rightarrow \, 0} \left[ 
Z^{\mbox{\footnotesize{RI$^\prime$}}}_\psi  
Z^{\mbox{\footnotesize{RI$^\prime$}}}_{{\cal O}}
\Sigma^{(T)}_{{\cal O}}(p) \right] \right|_{p^2 \, = \, \mu^2} ~=~ {\cal T}
\label{riopdef}
\end{equation}  
where $\Sigma^{(T)}_{{\cal O}}(p)$ is the tree part of $\langle \psi(-p) 
{\cal O}(0) \bar{\psi}(p) \rangle$ and ${\cal T}$ is the value of the tree term 
amplitude which may not necessarily be unity. In other words there is no $a$
dependence in ${\cal T}$. The explicit details of our choice of how to 
construct the Green's functions in terms of a basis of Lorentz tensors 
satisfying the same symmetry properties as the original operator is given in 
appendix A. This summarizes the procedure we will use to extract the 
renormalization constants of the operators which will then be encoded in the 
associated anomalous dimensions through the respective definitions 
\begin{equation}
\gamma^{\mbox{\footnotesize{$\MSbar$}}}_{{\cal O}} (a) ~=~ -~ \beta(a) 
\frac{\partial \ln Z^{\mbox{\footnotesize{$\MSbar$}}}_{{\cal O}}}  
{\partial a} ~-~ \alpha \gamma^{\mbox{\footnotesize{$\MSbar$}}}_\alpha(a) 
\frac{\partial \ln Z^{\mbox{\footnotesize{$\MSbar$}}}_{{\cal O}}}  
{\partial \alpha} 
\end{equation}
and 
\begin{equation}
\gamma^{\mbox{\footnotesize{RI$^\prime$}}}_{{\cal O}} (a) ~=~ -~ \beta(a) 
\frac{\partial \ln Z^{\mbox{\footnotesize{RI$^\prime$}}}_{{\cal O}}}
{\partial a} ~-~ \alpha \gamma^{\mbox{\footnotesize{RI$^\prime$}}}_\alpha(a) 
\frac{\partial \ln Z^{\mbox{\footnotesize{RI$^\prime$}}}_{{\cal O}}} 
{\partial \alpha} 
\end{equation}
where $\gamma^{\mbox{\footnotesize{RI$^\prime$}}}_\alpha(a)$ is given in 
\cite{22}.  

\sect{Anomalous dimensions.}
Having described in detail the method of renormalizing in the RI$^\prime$ 
scheme, we now record the explicit three loop results for the anomalous 
dimensions in this section. In constructing our results we made extensive use 
of the {\sc Mincer} package, \cite{34,35}, written in the symbolic manipulation
language {\sc Form}, \cite{36}. The {\sc Mincer} algorithm, \cite{34},
determines the divergent and finite parts of massless $2$-point functions using
dimensional regularization in $d$-dimensions. Therefore, it is ideal for the 
current problem since the Green's functions we are interested in are massless 
quark $2$-point functions with the appropriate operator inserted at zero 
momentum. This is the momentum configuration which is measured on the lattice. 
Moreover, since we are concerned with operators which are symmetrized in their 
Lorentz indices and satisfy various tracelessness conditions in addition to 
being flavour non-singlet operators, there is no possibility of mixing into 
other operators. This is an important observation since ordinarily nullifying 
the momentum flow through the operator could lead to the inability to correctly
determine the projection into the full basis of operators. (For a clear 
exposition on the deeper perils of operator mixing see, for example, 
\cite{37}.) The fact that each of the operators is multiplicatively 
renormalizable avoids this potential technicality. For our three loop 
computation we generated the Feynman diagrams with the {\sc Qgraf} package, 
\cite{38}. Specifically there are $3$ one loop, $37$ two loop and $684$ three 
loop diagrams to be calculated. Though for the operators with no three gluon 
and two quark leg operator insertions the latter total is reduced by $14$. 
Finally, the electronic {\sc Qgraf} output is converted into {\sc Form} input 
notation and the {\sc Form} version of the {\sc Mincer} algorithm, \cite{35}, 
is applied to the $724$ Feynman diagrams. The actual Feynman rules for each 
operator were generated automatically in {\sc Form}. First we constructed the 
object with the same symmetry and traceless conditions as the operators we are 
interested in. The explicit details for each operator are given in appendix B.
Then we applied an algorithm which systematically substitutes for the covariant
derivatives and functionally differentiates with respect to the various fields 
present to produce electronic forms of the $2$, $3$, $4$ and $5$-point operator
vertex insertions.   

Now we provide our results in $\MSbar$. For completeness we give those for the 
two Wilson operators and note that we found exact agreement with the results
first deduced in \cite{1,5,6,39,40}. These are 
\begin{eqnarray}  
\gamma^{\mbox{\footnotesize{$\MSbar$}}}_{\bar{\psi} \gamma^\mu D^\nu D^\sigma
\psi}(a) &=& \frac{25}{6} C_F a ~+~ C_F \left[ \frac{535}{27} C_A 
- \frac{2035}{432} C_F - \frac{415}{54} T_F \Nf \right] a^2 \nonumber \\
&& +~ C_F \left[ \left( \frac{55}{3} \zeta(3) + \frac{889433}{7776} \right) 
C_A^2 - \left( 55 \zeta(3) + \frac{311213}{15552} \right) C_A C_F \right. 
\nonumber \\
&& \left. ~~~~~~~~~~~-~ \left( \frac{200}{3} \zeta(3) + \frac{62249}{1944}
\right) C_A T_F \Nf + \left( \frac{110}{3} \zeta(3) - \frac{244505}{15552}
\right) C_F^2 \right. \nonumber \\
&& \left. ~~~~~~~~~~~+~ \left( \frac{200}{3} \zeta(3) - \frac{203627}{3888} 
\right) C_F T_F \Nf - \frac{2569}{486} T_F^2 \Nf^2 \right] a^3 \nonumber \\
&& +~ O(a^4) 
\end{eqnarray}  
and 
\begin{eqnarray}  
\gamma^{\mbox{\footnotesize{$\MSbar$}}}_{\bar{\psi} \gamma^\mu D^\nu D^\sigma
D^\rho \psi}(a) &=& \frac{157}{30} C_F a ~+~ \left[ 1292560 C_A 
- 287303 C_F - 530840 T_F \Nf \right] \frac{C_F a^2}{54000} \nonumber \\ 
&& +~ \left[ \left( 932472000 \zeta(3) + 6803318650 \right) C_A^2 \right.
\nonumber \\
&& \left. ~~~~~-~ \left( 2797416000 \zeta(3) + 1335140785 \right) C_A C_F 
\right. \nonumber \\
&& \left. ~~~~~-~ \left( 4069440000 \zeta(3) + 1760516200 \right) C_A T_F \Nf  
\right. \nonumber \\
&& \left. ~~~~~+~ \left( 1864944000 \zeta(3) - 714245693 \right) C_F^2 
\right. \nonumber \\
&& \left. ~~~~~+~ \left( 4069440000 \zeta(3) - 3304751260 \right) C_F T_F \Nf 
\right. \nonumber \\
&& \left. ~~~~~-~ 307421600 T_F^2 \Nf^2 \right] \frac{C_F a^3}{48600000} ~+~ 
O(a^4) 
\end{eqnarray}  
where we note that throughout the article when the operator appears explicitly
as a subscript on an object, it is regarded as a label and the free indices do
not endow the object with tensor properties. Likewise when we indicate the 
renormalization scheme on a quantity which is evaluated in perturbation theory,
that means that the variables in which it is expressed, such as $a$ and 
$\alpha$, are regarded as the variables in the {\em same} scheme. The 
relationship between the variables in either scheme is given in (\ref{cccon}) 
and (\ref{alpcon}). For the two transversity operators the $\MSbar$ anomalous 
dimensions have not been given previously and we find that  
\begin{eqnarray}  
\gamma^{\mbox{\footnotesize{$\MSbar$}}}_{\bar{\psi} \sigma^{\mu\nu} D^\sigma
D^\rho \psi}(a) &=& \frac{13}{3} C_F a ~+~ \left[ 1195 C_A - 311 C_F 
- 452 T_F \Nf \right] \frac{C_F a^2}{54} \nonumber \\ 
&& +~ \left[ \left( 10368 \zeta(3) + 126557 \right) C_A^2 - \left( 
31104 \zeta(3) + 30197 \right) C_A C_F \right. \nonumber \\
&& \left. ~~~~~-~ \left( 67392 \zeta(3) + 38900 \right) C_A T_F \Nf  
+ \left( 67392 \zeta(3) - 50552 \right) C_F T_F \Nf \right. \nonumber \\
&& \left. ~~~~~+~ \left( 20736 \zeta(3) - 17434 \right) C_F^2 
- 4816 T_F^2 \Nf^2 \right] \frac{C_F a^3}{972} ~+~ O(a^4) 
\label{tra3ms} 
\end{eqnarray}  
and 
\begin{eqnarray}  
\gamma^{\mbox{\footnotesize{$\MSbar$}}}_{\bar{\psi} \sigma^{\mu\nu} D^\sigma
D^\rho D^\lambda \psi}(a) &=& \frac{16}{3} C_F a ~+~ \left[ 1357 C_A 
- 296 C_F - 554 T_F \Nf \right] \frac{C_F a^2}{54} \nonumber \\ 
&& +~ \left[ \left( 272160 \zeta(3) + 2893009 \right) C_A^2 ~-~ \left( 816480 
\zeta(3) + 662155 \right) C_A C_F \right. \nonumber \\
&& \left. ~~~~~-~ \left( 1658880 \zeta(3) + 798892 \right) C_A T_F \Nf
\right. \nonumber \\
&& \left. ~~~~~+~ \left( 544320 \zeta(3) - 235100 \right) C_F^2 
\right. \nonumber \\
&& \left. ~~~~~+~ \left( 1658880 \zeta(3) - 1328860 \right) C_F T_F \Nf 
\right. \nonumber \\
&& \left. ~~~~~-~ 117776 T_F^2 \Nf^2 \right] \frac{C_F a^3}{19440} ~+~ 
O(a^4) ~.  
\label{tra4ms} 
\end{eqnarray}  
There are several checks on these two expressions. First, as we have computed
them in an arbitrary covariant gauge their final form must be independent of
the gauge parameter in a mass independent renormalization scheme, which is
apparent in (\ref{tra3ms}) and (\ref{tra4ms}), \cite{32,41}. Second, part of 
each of the three loop terms has in fact already been determined by the large 
$\Nf$ expansion in \cite{23}. There the leading order critical exponent 
corresponding to the anomalous dimension evaluated at the non-trivial 
$d$-dimensional fixed point of the QCD $\beta$-function was determined in 
$d$-dimensions using a method that was originally developed to study the 
perturbative structure of scalar field theories, \cite{42,43}. This critical 
exponent, \cite{23}, encodes all orders information on the corresponding 
renormalization group function at $O(1/\Nf)$. Therefore, if we formally write 
the leading $O(1/\Nf)$ part of the arbitrary $n$ transversity operator 
anomalous dimension as, \cite{23},  
\begin{equation} 
\gamma^{(n) \, \mbox{\footnotesize{$\MSbar$}}}_{\bar{\psi} \sigma^{\mu\nu_1} 
D^{\nu_2} \ldots D^{\nu_n} \psi}(a) ~=~ C_F \left[ b_1 a ~+~ 
\left( b_{21} T_F \Nf + b_{20} \right) a^2 ~+~ \sum_{r=3}^\infty 
\sum_{j=0}^{r-1} b_{rj} T_F^j \Nf^j a^r \right] 
\end{equation} 
then the leading order coefficient of the $\Nf$ polynomial at three loops is
given by 
\begin{equation}
b_{32} ~=~ \frac{4}{27} \left[ 48 S_3(n) - 80 S_2(n) - 16 S_1(n) 
+ \frac{3( 17 n^2 + 17 n - 8 )}{n(n+1)} \right]  
\end{equation}
where $S_l(n)$~$=$~$\sum_{i=1}^n 1/i^l$. Evaluating this for $n$~$=$~$3$ and
$n$~$=$~$4$ reproduces the corresponding coefficients in (\ref{tra3ms}) and
(\ref{tra4ms}) respectively. As a final check we note that we have used the
method of \cite{44} to perform the automatic renormalization of Green's 
functions at high loop order. This entails computing the Green's functions as a
function of the bare coupling constant and bare gauge parameter. The 
counterterms are then introduced at the end of the computation by rescaling by 
the known renormalization constants. Therefore, the remaining divergence is 
absorbed by the renormalization constant associated with the Green's function. 
Moreover, given the way it has been deduced, the non-simple poles in $\epsilon$
are constrained to satisfy a specific form depending on the lower order simple
poles due to the underlying renormalization group equation. This is a 
non-trivial checking criterion, especially in the presence of parameters such 
as the gauge parameter and group Casimirs, and it is reassuring to record that 
all the renormalization constants determined for the above anomalous dimensions
precisely satisfied this criterion. Implicit in this final check is the fact
that the already known two loop anomalous dimensions of \cite{27,28,29,30,31}
are correctly reproduced when the $n$-dependent results are evaluated for 
$n$~$=$~$3$ and $n$~$=$~$4$. All these checks therefore give us confidence that
not only are all our expressions correct but also, for example, that the 
original Feynman rules were correctly generated.  

Having established the $\MSbar$ anomalous dimensions it is then relatively
straightforward to deduce the anomalous dimensions in the RI$^\prime$ scheme.
This is achieved by replacing the renormalization constants which scale the
bare internal parameters and that of the overall quark wave function, by those 
of the RI$^\prime$ scheme and then impose the RI$^\prime$ scheme definition for
the operator renormalization, (\ref{riopdef}). As a check on the resulting 
renormalization constants, the non-simple poles in $\epsilon$ also have to 
satisfy various constraints emanating from the renormalization group equation, 
similar to those of the $\MSbar$ scheme. We note, for completeness, that these 
are fulfilled. Therefore, we record the corresponding three loop RI$^\prime$ 
anomalous dimensions are, in four dimensions,  
\begin{eqnarray}  
\gamma^{\mbox{\footnotesize{RI$^\prime$}}}_{\bar{\psi} \gamma^\mu D^\nu 
D^\sigma \psi}(a) &=& \frac{25}{6} C_F a ~+~ \left[ \left( 324 \alpha^2 + 972 
\alpha + 17976 \right) C_A - 2035 C_F - 6744 T_F \Nf \right] 
\frac{C_F a^2}{432} \nonumber \\ 
&& +~ \left[ \left( 29160 \alpha^4 + 260820 \alpha^3 - 69984 \zeta(3) \alpha^2 
+ 1257768 \alpha^2 \right. \right. \nonumber \\
&& \left. \left. ~~~~~~-~ 723168 \zeta(3) \alpha + 4103676 \alpha 
- 6443712 \zeta(3) + 50460154 \right) C_A^2 \right. \nonumber \\
&& \left. ~~~~~+~ \left( 3240 \alpha^3 - 91260 \alpha^2 - 1043460 \alpha
- 171072 \zeta(3) - 8028146 \right) C_A C_F \right. \nonumber \\
&& \left. ~~~~~-~ \left( 259200 \alpha^2 - 186624 \zeta(3) \alpha 
+ 1401408 \alpha + 2322432 \zeta(3) \right. \right. \nonumber \\ 
&& \left. \left. ~~~~~~~~~~~~+~ 35016976 \right) C_A T_F \Nf \right. 
\nonumber \\ 
&& \left. ~~~~~+~ \left( 269280 \alpha + 3691008 \zeta(3) - 3568016 
\right) C_F T_F \Nf \right. \nonumber \\
&& \left. ~~~~~+~ \left( 2851200 \zeta(3) - 1222525 \right) C_F^2 
+ 5492800 T_F^2 \Nf^2 \right] \frac{C_F a^3}{77760} \nonumber \\
&& +~ O(a^4)
\end{eqnarray}  
and 
\begin{eqnarray}  
\gamma^{\mbox{\footnotesize{RI$^\prime$}}}_{\bar{\psi} \sigma^{\mu\nu} D^\sigma
D^\rho \psi}(a) &=& \frac{13}{3} C_F a ~+~ \left[ \left( 99 \alpha^2 + 297 
\alpha + 4788 \right) C_A - 622 C_F - 1776 T_F \Nf \right] 
\frac{C_F a^2}{108} \nonumber \\ 
&& +~ \left[ \left( 8910 \alpha^4 + 80055 \alpha^3 - 23328 \zeta(3) \alpha^2 
+ 387846 \alpha^2 \right. \right. \nonumber \\
&& \left. \left. ~~~~~~-~ 241056 \zeta(3) \alpha + 1279197 \alpha 
- 1902528 \zeta(3) + 13940156 \right) C_A^2 \right. \nonumber \\
&& \left. ~~~~~+~ \left( 6210 \alpha^3 + 3420 \alpha^2 - 157170 \alpha
+ 746496 \zeta(3) - 2737412 \right) C_A C_F \right. \nonumber \\
&& \left. ~~~~~-~ \left( 79200 \alpha^2 - 62208 \zeta(3) \alpha 
+ 434736 \alpha + 580608 \zeta(3) \right. \right. \nonumber \\ 
&& \left. \left. ~~~~~~~~~~~~+~ 9592064 \right) C_A T_F \Nf \right. 
\nonumber \\ 
&& \left. ~~~~~+~ \left( 40560 \alpha + 850176 \zeta(3) - 706112 
\right) C_F T_F \Nf \right. \nonumber \\
&& \left. ~~~~~+~ \left( 414720 \zeta(3) - 348680 \right) C_F^2 
+ 1491200 T_F^2 \Nf^2 \right] \frac{C_F a^3}{19440} \nonumber \\
&& +~ O(a^4) 
\end{eqnarray}  
for $n$~$=$~$3$. Further, for the $n$~$=$~$4$ operators we have 
\begin{eqnarray}  
\gamma^{\mbox{\footnotesize{RI$^\prime$}}}_{\bar{\psi} \gamma^\mu D^\nu 
D^\sigma D^\rho \psi}(a) &=& \frac{157}{30} C_F a ~+~ \left[ \left( 
63000 \alpha^2 + 189000 \alpha + 2939040 \right) C_A - 287303 C_F 
\right. \nonumber \\
&& \left. ~~~~~~~~~~~~~~~~~~~-~ 1129560 T_F \Nf \right] \frac{C_F a^2}{54000} 
\nonumber \\ 
&& +~ \left[ \left( 28350000 \alpha^4 + 264937500 \alpha^3 - 87480000 \zeta(3) 
\alpha^2 \right. \right. \nonumber \\
&& \left. \left. ~~~~~~+~ 1290594375 \alpha^2 - 903960000 \zeta(3) \alpha 
+ 4337679375 \alpha \right. \right. \nonumber \\
&& \left. \left. ~~~~~~-~ 4769928000 \zeta(3) + 44592646550 \right) C_A^2 
\right. \nonumber \\
&& \left. ~~~~~+~ \left( 27675000 \alpha^3 + 58466250 \alpha^2 
- 253773750 \alpha \right. \right. \nonumber \\
&& \left. \left. ~~~~~~~~~~~~+~ 624024000 \zeta(3) - 8248198970 \right) C_A 
C_F \right. \nonumber \\
&& \left. ~~~~~-~ \left( 252000000 \alpha^2 - 233280000 \zeta(3) \alpha 
+ 1452285000 \alpha \right. \right. \nonumber \\ 
&& \left. \left. ~~~~~~~~~~~~+~ 1995840000 \zeta(3) + 31068538400 \right) C_A 
T_F \Nf \right. \nonumber \\ 
&& \left. ~~~~~+~ \left( 65490000 \alpha + 2825280000 \zeta(3) - 2407455920 
\right) C_F T_F \Nf \right. \nonumber \\
&& \left. ~~~~~+~ \left( 1864944000 \zeta(3) - 714245693 \right) C_F^2 
\right. \nonumber \\
&& \left. ~~~~~+~ 4979028800 T_F^2 \Nf^2 \right] \frac{C_F a^3}{48600000} ~+~ 
O(a^4) 
\end{eqnarray}  
and 
\begin{eqnarray}  
\gamma^{\mbox{\footnotesize{RI$^\prime$}}}_{\bar{\psi} \sigma^{\mu\nu} D^\sigma
D^\rho D^\lambda \psi}(a) &=& \frac{16}{3} C_F a ~+~ \left[ \left( 
225 \alpha^2 + 675 \alpha + 11874 \right) C_A - 1184 C_F - 4560 T_F \Nf \right] 
\frac{C_F a^2}{216} \nonumber \\ 
&& +~ \left[ \left( 16200 \alpha^4 + 150768 \alpha^3 - 46656 \zeta(3) \alpha^2 
+ 731421 \alpha^2 \right. \right. \nonumber \\
&& \left. \left. ~~~~~~-~ 482112 \zeta(3) \alpha + 2435409 \alpha 
- 3214080 \zeta(3) + 27763364 \right) C_A^2 \right. \nonumber \\
&& \left. ~~~~~+~ \left( 8964 \alpha^3 - 8244 \alpha^2 - 363072 \alpha
+ 518400 \zeta(3) - 5050688 \right) C_A C_F \right. \nonumber \\
&& \left. ~~~~~-~ \left( 144000 \alpha^2 - 124416 \zeta(3) \alpha 
+ 818712 \alpha + 1327104 \zeta(3) \right. \right. \nonumber \\ 
&& \left. \left. ~~~~~~~~~~~~+~ 19484432 \right) C_A T_F \Nf \right. 
\nonumber \\ 
&& \left. ~~~~~+~ \left( 93696 \alpha + 1990656 \zeta(3) - 1687424 
\right) C_F T_F \Nf \right. \nonumber \\
&& \left. ~~~~~+~ \left( 870912 \zeta(3) - 376160 \right) C_F^2 
+ 3138560 T_F^2 \Nf^2 \right] \frac{C_F a^3}{31104} \nonumber \\
&& +~ O(a^4) 
\end{eqnarray}  
in four dimensions. Clearly they all satisfy the trivial check that the one
loop term is scheme independent. Though since the RI$^\prime$ scheme is a mass
dependent one, the anomalous dimensions will not necessarily be independent of
the gauge parameter as is clearly the case above. 

Although we have performed the computation in an arbitrary gauge and colour
group, for practical purposes it is useful to specify the results for $SU(3)$.
Therefore, the $\MSbar$ transversity anomalous dimensions are 
\begin{eqnarray} 
\left. \gamma^{\mbox{\footnotesize{$\MSbar$}}}_{\bar{\psi} \sigma^{\mu\nu}
D^\sigma D^\rho \psi}(a) \right|^{SU(3)} &=& \frac{52}{9} a ~-~ 
2\left[ 678 \Nf - 9511 \right] \frac{a^2}{243} \nonumber \\
&& -~ \left[ 10836 \Nf^2 + 505440 \zeta(3) \Nf + 828462 \Nf \right. 
\nonumber \\
&& \left. ~~~~~~-~ 51840 \zeta(3) - 8885081 \right] \frac{a^3}{6561} ~+~ O(a^4) 
\end{eqnarray}  
and 
\begin{eqnarray} 
\left. \gamma^{\mbox{\footnotesize{$\MSbar$}}}_{\bar{\psi} \sigma^{\mu\nu}
D^\sigma D^\rho D^\lambda \psi}(a) \right|^{SU(3)} &=& \frac{64}{9} a ~-~ 
2\left[ 831 \Nf - 11029 \right] \frac{a^2}{243} \nonumber \\
&& -~ \left[ 264996 \Nf^2 + 12441600 \zeta(3) \Nf + 18758202 \Nf \right. 
\nonumber \\
&& \left. ~~~~~~-~ 1360800 \zeta(3) - 206734549 \right] \frac{a^3}{131220} ~+~ 
O(a^4) 
\end{eqnarray}  
where $T_F$~$=$~$1/2$, $C_F$~$=$~$4/3$ and $C_A$~$=$~$3$ for $SU(3)$. In
addition for the RI$^\prime$ scheme we record each of the anomalous dimensions
in the Landau gauge since that is the gauge primarily used in matching to 
lattice results. We have 
\begin{eqnarray} 
\left. \gamma^{\mbox{\footnotesize{RI$^\prime$}}}_{\bar{\psi} \gamma^\mu D^\nu 
D^\sigma \psi}(a) \right|_{\alpha = 0}^{SU(3)} &=& \frac{50}{9} a ~-~ 
\left[ 2529 \Nf - 38411 \right] \frac{a^2}{243} \nonumber \\
&& +~ \left[ 6179400 \Nf^2 - 4603392 \zeta(3) \Nf - 247068636 \Nf \right. 
\nonumber \\
&& \left. ~~~~~-~ 241240032 \zeta(3) + 1889349409 \right] 
\frac{a^3}{262440} ~+~ O(a^4) 
\end{eqnarray}  
and 
\begin{eqnarray} 
\left. \gamma^{\mbox{\footnotesize{RI$^\prime$}}}_{\bar{\psi} \gamma^\mu D^\nu
D^\sigma D^\rho \psi}(a) \right|_{\alpha = 0}^{SU(3)} &=& \frac{314}{45} a ~-~ 
\left[ 423585 \Nf - 6325537 \right] \frac{a^2}{30375} \nonumber \\
&& +~ \left[ 5601407400 \Nf^2 - 4996080000 \zeta(3) \Nf - 216935001960 \Nf 
\right. \nonumber \\
&& \left. ~~~~~-~ 167030100000 \zeta(3) + 1651820638271 \right] 
\frac{a^3}{164025000} \nonumber \\
&& +~ O(a^4) 
\end{eqnarray}  
for the Wilson operators. Whilst 
\begin{eqnarray} 
\left. \gamma^{\mbox{\footnotesize{RI$^\prime$}}}_{\bar{\psi} \sigma^{\mu\nu}
D^\sigma D^\rho \psi}(a) \right|_{\alpha = 0}^{SU(3)} &=& \frac{52}{9} a ~-~ 
4\left[ 666 \Nf - 10151 \right] \frac{a^2}{243} \nonumber \\
&& +~ \left[ 838800 \Nf^2 - 684288 \zeta(3) \Nf - 33432384 \Nf \right. 
\nonumber \\
&& \left. ~~~~~-~ 30148848 \zeta(3) + 256256731 \right] \frac{a^3}{32805} ~+~ 
O(a^4) 
\end{eqnarray}  
and 
\begin{eqnarray} 
\left. \gamma^{\mbox{\footnotesize{RI$^\prime$}}}_{\bar{\psi} \sigma^{\mu\nu}
D^\sigma D^\rho D^\lambda \psi}(a) \right|_{\alpha = 0}^{SU(3)} &=& 
\frac{64}{9} a ~-~ 5 \left[ 684 \Nf - 10213 \right] \frac{a^2}{243} 
\nonumber \\
&& +~ \left[ 1765440 \Nf^2 - 1492992 \zeta(3) \Nf - 68291094 \Nf 
\right. \nonumber \\
&& \left. ~~~~~-~ 56935872 \zeta(3) + 515247289 \right] 
\frac{a^3}{52488} ~+~ O(a^4) 
\end{eqnarray}  
for the transversity case. 

\sect{Finite parts.} 
In this section we record the three loop $\MSbar$ and RI$^\prime$ expressions 
for the amplitudes of the various Green's functions we computed to obtain the 
previous anomalous dimensions. These are essential for lattice matching 
computations which therefore necessitates their tedious presentation. The 
specific definitions of the quantities $\Sigma^{(i)}_{{\cal O}}(p)$ are, as
noted before, given in appendix A. It is worth pointing out that not all the 
amplitudes have an $a$ independent leading term.  

First, for the Wilson operator with $n$~$=$~$3$, we have 
\begin{eqnarray} 
\left. \Sigma^{(1) ~ {\mbox{\footnotesize{$\MSbar$}}} ~ 
\mbox{\footnotesize{finite}}}_{\bar{\psi} \gamma^\mu 
D^\nu D^\sigma \psi}(p) \right|_{p^2 \, = \, \mu^2} &=& \left( \frac{1}{3}
+ \frac{2}{3} \alpha \right) C_F a \nonumber \\
&& +~ \left[ \left( \frac{367}{30} - \frac{6}{5} \zeta(3) \alpha 
+ \frac{361}{90} \alpha + \frac{7}{9} \alpha^2 - \frac{6}{5} \zeta(3) \right)
C_F C_A
\right. \nonumber \\
&& \left. ~~~~~ 
+~ \left( -~ \frac{1087}{120} - \frac{37}{18} \alpha + \frac{1}{9} \alpha^2
+ \frac{24}{5} \zeta(3) \right) C_F^2 - \frac{25}{9} \Nf T_F C_F 
\right] a^2 \nonumber \\
&& + \left[ \left( -~ \frac{36151}{243} + \frac{112}{15} \zeta(3) \alpha 
- \frac{14671}{810} \alpha + \frac{154}{9} \zeta(3) \right) \Nf T_F C_F C_A 
\right. \nonumber \\
&& \left. ~~~~~
+~ \left( \frac{504013}{6480} + \frac{22735}{1944} \alpha 
- \frac{224}{5} \zeta(3) \right) \Nf T_F C_F^2 
+ \frac{4210}{243} \Nf^2 T_F^2 C_F  
\right. \nonumber \\
&& \left. ~~~~~
+~ \left( \frac{6480923}{19440} - \frac{3727}{120} \zeta(3) \alpha 
+ 2 \zeta(5) \alpha + \frac{759413}{12960} \alpha
\right. \right. \nonumber \\
&& \left. \left. ~~~~~~~~~~~
-~ \frac{47}{15} \zeta(3) \alpha^2 + \frac{1}{6} \zeta(5) \alpha^2 
+ \frac{49223}{4320} \alpha^2 + \frac{401}{216} \alpha^3
\right. \right. \nonumber \\
&& \left. \left. ~~~~~~~~~~~
-~ \frac{2563}{24} \zeta(3) - \frac{67}{2} \zeta(5) \right) C_F C_A^2 
\right. \nonumber \\
&& \left. ~~~~~
+~ \left( -~ \frac{2811619}{6480} + \frac{25}{3} \zeta(3) \alpha 
+ 8 \zeta(5) \alpha - \frac{51839}{1215} \alpha - \zeta(3) \alpha^2 
\right. \right. \nonumber \\
&& \left. \left. ~~~~~~~~~~~
-~ \frac{613}{162} \alpha^2 - \frac{1}{6} \alpha^3 + \frac{979}{15} \zeta(3)
+ \frac{784}{3} \zeta(5) \right) C_F^2 C_A 
\right. \nonumber \\
&& \left. ~~~~~
+~ \left( \frac{28855943}{155520} - 4 \zeta(3) \alpha 
+ \frac{218971}{15552} \alpha + \frac{539}{162} \alpha^2
\right. \right. \nonumber \\
&& \left. \left. ~~~~~~~~~~~
-~ \frac{11}{54} \alpha^3 + \frac{860}{9} \zeta(3) - 272 \zeta(5) \right) C_F^3 
\right] a^3 ~+~ O(a^4)  
\end{eqnarray} 
and 
\begin{eqnarray} 
\left. \Sigma^{(2) ~ {\mbox{\footnotesize{$\MSbar$}}} ~ 
\mbox{\footnotesize{finite}}}_{\bar{\psi} \gamma^\mu 
D^\nu D^\sigma \psi}(p) \right|_{p^2 \, = \, \mu^2} &=& -~ \frac{1}{3} ~+~ 
\left( \frac{107}{54} + \frac{1}{6} \alpha \right) C_F a \nonumber \\  
&& +~ \left[ \left( \frac{86597}{4860} + \frac{2}{5} \zeta(3) \alpha 
+ \frac{167}{360} \alpha + \frac{13}{72} \alpha^2 - \frac{18}{5} \zeta(3)
\right) C_F C_A 
\right. \nonumber \\
&& \left. ~~~~~ 
+~ \left( -~ \frac{1471891}{155520} - \frac{401}{216} \alpha
+ \frac{5}{36} \alpha^2 + \frac{12}{5} \zeta(3) \right) C_F^2 
\right. \nonumber \\
&& \left. ~~~~~ 
-~ \frac{32363}{3888} \Nf T_F C_F  
\right] a^2 \nonumber \\
&& +~ \left[ \left( -~ \frac{30365437}{209952}
+ \frac{68}{45} \zeta(3) \alpha - \frac{1474}{405} \alpha
\right. \right. \nonumber \\
&& \left. \left. ~~~~~~~
-~ \frac{3577}{243} \zeta(3) - \frac{100}{9} \zeta(4) \right) \Nf T_F C_F C_A   
\right. \nonumber \\
&& \left. ~~~~~
+~ \left( \frac{1019471}{2099520} - \frac{100}{27} \zeta(3) \alpha 
+ \frac{26413}{2592} \alpha
\right. \right. \nonumber \\
&& \left. \left. ~~~~~~~~~~
+~ \frac{4166}{135} \zeta(3) + \frac{100}{9} \zeta(4) \right) \Nf T_F C_F^2
\right. \nonumber \\
&& \left. ~~~~~
+~ \left( \frac{1227463}{52488} + \frac{400}{243} \zeta(3) \right)
\Nf^2 T_F^2 C_F
\right. \nonumber \\
&& \left. ~~~~~
+~ \left( \frac{208545851}{1049760} - \frac{3721}{720} \zeta(3) \alpha 
- \frac{1}{8} \zeta(4) \alpha + \frac{1}{6} \zeta(5) \alpha 
\right. \right. \nonumber \\
&& \left. \left. ~~~~~~~~~~~
+~ \frac{390323}{25920} \alpha
- \frac{23}{720} \zeta(3) \alpha^2 - \frac{1}{16} \zeta(4) \alpha^2 
\right. \right. \nonumber \\
&& \left. \left. ~~~~~~~~~~~
-~ \frac{17}{36} \zeta(5) \alpha^2 
+ \frac{30983}{8640} \alpha^2 + \frac{1}{9} \zeta(3) \alpha^3 
+ \frac{475}{864} \alpha^3 
\right. \right. \nonumber \\
&& \left. \left. ~~~~~~~~~~~
-~ \frac{31163}{972} \zeta(3) 
+ \frac{647}{144} \zeta(4) + \frac{13}{4} \zeta(5) \right) C_F C_A^2
\right. \nonumber \\
&& \left. ~~~~~
+~ \left( -~ \frac{781692217}{8398080} + \frac{697}{54} \zeta(3) \alpha 
- \frac{8}{3} \zeta(5) \alpha - \frac{57691}{1620} \alpha
\right. \right. \nonumber \\
&& \left. \left. ~~~~~~~~~~~
+~ \frac{7}{18} \zeta(3) \alpha^2 - \frac{7895}{2592} \alpha^2
- \frac{1}{3} \zeta(3) \alpha^3 + \frac{1}{24} \alpha^3 
- \frac{743}{540} \zeta(3)
\right. \right. \nonumber \\
&& \left. \left. ~~~~~~~~~~~
-~ \frac{67}{6} \zeta(4) - \frac{4}{9} \zeta(5) \right) C_F^2 C_A 
\right. \nonumber \\
&& \left. ~~~~~
+~ \left( -~ \frac{1161367}{43740} - \frac{733}{54} \zeta(3) \alpha 
+ \frac{458737}{20736} \alpha - \frac{25}{9} \zeta(3) \alpha^2 
\right. \right. \nonumber \\
&& \left. \left. ~~~~~~~~~~~
+~ \frac{617}{324} \alpha^2
+ \frac{2}{9} \zeta(3) \alpha^3 - \frac{31}{216} \alpha^3
+ \frac{9203}{972} \zeta(3)
\right. \right. \nonumber \\
&& \left. \left. ~~~~~~~~~~~
+~ \frac{55}{9} \zeta(4) + 24 \zeta(5) \right) C_F^3 \right] a^3 ~+~ O(a^4) ~. 
\end{eqnarray} 
For the $n$~$=$~$4$ case we have 
\begin{eqnarray} 
\left. \Sigma^{(1) ~ {\mbox{\footnotesize{$\MSbar$}}} ~ 
\mbox{\footnotesize{finite}}}_{\bar{\psi} \gamma^\mu 
D^\nu D^\sigma D^\rho \psi}(p) \right|_{p^2 \, = \, \mu^2} &=& -~ 1
+ \left( \frac{1871}{225} + \frac{4}{3} \alpha \right) C_F a \nonumber \\ 
&& +~ \left[ \left( \frac{26869109}{324000} - \frac{3}{5} \zeta(3) \alpha 
+ \frac{10313}{1440} \alpha + \frac{245}{144} \alpha^2 - 13 \zeta(3) \right) 
C_F C_A 
\right. \nonumber \\
&& \left. ~~~~~ 
+~ \left( -~ \frac{345682991}{6480000} - \frac{43553}{3600} \alpha
+ \frac{13}{72} \alpha^2 + \frac{48}{5} \zeta(3) \right) C_F^2 
\right. \nonumber \\
&& \left. ~~~~~
-~ \frac{6041063}{162000} \Nf T_F C_F \right] a^2 \nonumber \\
&& +~ \left[  \left( -~ \frac{32796795659}{43740000}
+ \frac{46}{3} \zeta(3) \alpha - \frac{48917}{1296} \alpha
\right. \right. \nonumber \\
&& \left. \left. ~~~~~~
-~ \frac{139334}{2025} \zeta(3) - \frac{628}{15} \zeta(4) \right) 
\Nf T_F C_F C_A 
\right. \nonumber \\
&& \left. ~~~~~
+~ \left( \frac{63233459093}{437400000} - \frac{628}{45} \zeta(3) \alpha 
+ \frac{2010352}{30375} \alpha
\right. \right. \nonumber \\
&& \left. \left. ~~~~~~~~~~~~
+~ \frac{77018}{675} \zeta(3) + \frac{628}{15} \zeta(4) \right) \Nf T_F C_F^2  
\right. \nonumber \\
&& \left. ~~~~~
+~ \left( \frac{1335574847}{10935000} + \frac{2512}{405} \zeta(3) \right) 
\Nf^2 T_F^2 C_F 
\right. \nonumber \\
&& \left. ~~~~~
+~ \left( \frac{1578785326351}{1399680000} - \frac{43861}{720} \zeta(3) \alpha 
- \frac{3}{8} \zeta(4) \alpha + \frac{5}{2} \zeta(5) \alpha 
\right. \right. \nonumber \\
&& \left. \left. ~~~~~~~~~~~~
+~ \frac{56000717}{414720} \alpha - \frac{1753}{360} \zeta(3) \alpha^2 
- \frac{3}{16} \zeta(4) \alpha^2 
\right. \right. \nonumber \\
&& \left. \left. ~~~~~~~~~~~~
-~ \frac{4}{3} \zeta(5) \alpha^2 + \frac{986237}{34560} \alpha^2
+ \frac{1}{3} \zeta(3) \alpha^3 + \frac{7859}{1728} \alpha^3
\right. \right. \nonumber \\
&& \left. \left. ~~~~~~~~~~~~
-~ \frac{2350679}{10125} \zeta(3) + \frac{16687}{1200} \zeta(4)
+ \frac{91}{3} \zeta(5) \right) C_F C_A^2 
\right. \nonumber \\
&& \left. ~~~~~
+~ \left( -~ \frac{1552257600373}{1749600000}
+ \frac{135041}{2250} \zeta(3) \alpha + 4 \zeta(5) \alpha 
\right. \right. \nonumber \\
&& \left. \left. ~~~~~~~~~~~~
-~ \frac{181148459}{777600} \alpha + \frac{32}{15} \zeta(3) \alpha^2 
- \frac{1336787}{51840} \alpha^2
\right. \right. \nonumber \\
&& \left. \left. ~~~~~~~~~~~~
-~ \zeta(3) \alpha^3 - \frac{205}{192} \alpha^3 
+ \frac{395693}{2700} \zeta(3)
\right. \right. \nonumber \\
&& \left. \left. ~~~~~~~~~~~~
-~ \frac{1739}{50} \zeta(4) + \frac{116}{3} \zeta(5) \right) C_F^2 C_A 
\right. \nonumber \\
&& \left. ~~~~~
+~ \left( \frac{203923883969}{2916000000} - \frac{29329}{450} \zeta(3) \alpha 
+ \frac{685201111}{4860000} \alpha
\right. \right. \nonumber \\
&& \left. \left. ~~~~~~~~~~~~
-~ \frac{157}{15} \zeta(3) \alpha^2 + \frac{911003}{64800} \alpha^2
+ \frac{2}{3} \zeta(3) \alpha^3 - \frac{565}{864} \alpha^3
\right. \right. \nonumber \\
&& \left. \left. ~~~~~~~~~~~~
-~ \frac{910609}{40500} \zeta(3) + \frac{1439}{75} \zeta(4) + 96 \zeta(5)
\right) C_F^3 \right] a^3 \nonumber \\
&& +~ O(a^4) 
\end{eqnarray} 
and 
\begin{eqnarray} 
\left. \Sigma^{(2) ~ {\mbox{\footnotesize{$\MSbar$}}} ~ 
\mbox{\footnotesize{finite}}}_{\bar{\psi} \gamma^\mu 
D^\nu D^\sigma D^\rho \psi}(p) \right|_{p^2 \, = \, \mu^2} &=& \left(
\frac{3}{160} + \frac{1}{32} \alpha \right) C_F a \nonumber \\  
&& +~ \left[ \left( \frac{70559}{115200} - \frac{3}{40} \zeta(3) \alpha 
+ \frac{4991}{23040} \alpha + \frac{31}{768} \alpha^2 - \frac{1}{40} \zeta(3)
\right) C_F C_A \right. \nonumber \\ 
&& \left. ~~~~~
+~ \left( -~ \frac{40907}{96000} - \frac{7453}{57600} \alpha
- \frac{1}{384} \alpha^2 + \frac{1}{5} \zeta(3) \right) C_F^2 
\right. \nonumber \\
&& \left. ~~~~~ 
-~ \frac{119}{800} \Nf T_F C_F \right] a^2  \nonumber \\ 
&& +~ \left[ \left( -~ \frac{6448567}{864000} + \frac{53}{120} \zeta(3) \alpha 
- \frac{104179}{103680} \alpha + \frac{53}{216} \zeta(3) \right)
\Nf T_F C_F C_A 
\right. \nonumber \\
&& \left. ~~~~~
+~ \left( \frac{28776619}{8640000} + \frac{4054403}{5184000} \alpha
- \frac{77}{54} \zeta(3) \right) \Nf T_F C_F^2 
\right. \nonumber \\
&& \left. ~~~~~
+~ \left( \frac{818118521}{55296000} - \frac{4003}{2160} \zeta(3) \alpha 
+ \frac{5}{48} \zeta(5) \alpha + \frac{22103567}{6635520} \alpha
\right. \right. \nonumber \\
&& \left. \left. ~~~~~~~~~~~~
-~ \frac{6641}{34560} \zeta(3) \alpha^2 + \frac{1}{96} \zeta(5) \alpha^2 
+ \frac{69635}{110592} \alpha^2 + \frac{913}{9216} \alpha^3
\right. \right. \nonumber \\
&& \left. \left. ~~~~~~~~~~~~
-~ \frac{198233}{28800} \zeta(3) + \frac{263}{96} \zeta(5) \right) C_F C_A^2 
\right. \nonumber \\
&& \left. ~~~~~
+~ \left( -~ \frac{318424073}{25920000} + \frac{7129}{12000} \zeta(3) \alpha 
+ \frac{1}{2} \zeta(5) \alpha 
- \frac{20896549}{6912000} \alpha
\right. \right. \nonumber \\
&& \left. \left. ~~~~~~~~~~~~
-~ \frac{1}{80} \zeta(3) \alpha^2 - \frac{502243}{1382400} \alpha^2
- \frac{29}{1024} \alpha^3 + \frac{1396697}{108000} \zeta(3)
\right. \right. \nonumber \\
&& \left. \left. ~~~~~~~~~~~~
-~ 8 \zeta(5) \right) C_F^2 C_A + \frac{51959}{54000} \Nf^2 T_F^2 C_F 
\right. \nonumber \\
&& \left. ~~~~~
+~ \left( \frac{37693459}{345600000} - \frac{13}{60} \zeta(3) \alpha 
+ \frac{239928131}{207360000} \alpha + \frac{27433}{115200} \alpha^2
\right. \right. \nonumber \\
&& \left. \left. ~~~~~~~~~~~~
-~ \frac{41}{4608} \alpha^3 - \frac{272419}{27000} \zeta(3)
+ \frac{38}{3} \zeta(5) \right) C_F^3 \right] a^3 \nonumber \\
&& +~ O(a^4) ~. 
\end{eqnarray}  
Turning to the case of the transversity operator, for $n$~$=$~$3$ we have 
\begin{eqnarray} 
\left. \Sigma^{(1) ~ {\mbox{\footnotesize{$\MSbar$}}} ~ 
\mbox{\footnotesize{finite}}}_{\bar{\psi} \sigma^{\mu\nu}  
D^\sigma D^\rho \psi}(p) \right|_{p^2 \, = \, \mu^2} &=& \frac{1}{18} \left[
1 ~+~ \left[ \left( -~ \frac{109}{18} - \frac{5}{6} \alpha \right) C_F \right] 
a \right. \nonumber \\ 
&& \left. +~ \left[ \left( -~ \frac{48941}{810} - \frac{3}{5} \zeta(3) \alpha 
- \frac{1223}{360} \alpha - \frac{67}{72} \alpha^2 + \frac{59}{5} \zeta(3)
\right) C_F C_A
\right. \right. \nonumber \\
&& \left. \left. ~~~~~ 
+\, \left( \frac{26467}{810} + \frac{119}{18} \alpha - \frac{17}{36} \alpha^2
- \frac{48}{5} \zeta(3) \! \right) C_F^2 + \frac{2197}{81} \Nf T_F C_F 
\right] a^2 \right. \nonumber \\
&& \left. +~ \left[ \left( \frac{2254181}{4374} - \frac{124}{15} \zeta(3) 
\alpha + \frac{32359}{1620} \alpha
\right. \right. \right. \nonumber \\
&& \left. \left. \left. ~~~~~~~~
+~ \frac{3404}{81} \zeta(3) + \frac{104}{3} \zeta(4) \right) \Nf T_F C_F C_A 
\right. \right. \nonumber \\
&& \left. \left. ~~~~~
+~ \left( -~ \frac{699751}{21870} + \frac{104}{9} \zeta(3) \alpha 
- \frac{18349}{486} \alpha
\right. \right. \right. \nonumber \\
&& \left. \left. \left. ~~~~~~~~~~~~
-~ \frac{11176}{135} \zeta(3) - \frac{104}{3} \zeta(4) \right) \Nf T_F C_F^2 
\right. \right. \nonumber \\
&& \left. \left. ~~~~~
+~ \left( -~ \frac{177970}{2187} - \frac{416}{81} \zeta(3) \right) 
\Nf^2 T_F^2 C_F 
\right. \right. \nonumber \\
&& \left. \left. ~~~~~
+~ \left( -~ \frac{260680459}{349920} + \frac{5537}{180} \zeta(3) \alpha 
+ \frac{3}{8} \zeta(4) \alpha - \frac{3}{2} \zeta(5) \alpha 
\right. \right. \right. \nonumber \\
&& \left. \left. \left. ~~~~~~~~~~~~
-~ \frac{95953}{1296} \alpha + \frac{133}{80} \zeta(3) \alpha^2 
+ \frac{3}{16} \zeta(4) \alpha^2 + \frac{4}{3} \zeta(5) \alpha^2 
\right. \right. \right. \nonumber \\
&& \left. \left. \left. ~~~~~~~~~~~~
-~ \frac{35543}{2160} \alpha^2 - \frac{1}{3} \zeta(3) \alpha^3 
- \frac{2227}{864} \alpha^3
\right. \right. \right. \nonumber \\
&& \left. \left. \left. ~~~~~~~~~~~~
+~ \frac{194707}{1296} \zeta(3) - \frac{463}{48} \zeta(4) 
- \frac{139}{6} \zeta(5) \right) C_F C_A^2 
\right. \right. \nonumber \\
&& \left. \left. ~~~~~
+~ \left( \frac{77900401}{174960} - \frac{3983}{90} \zeta(3) \alpha 
+ 4 \zeta(5) \alpha + \frac{635981}{4860} \alpha
\right. \right. \right. \nonumber \\
&& \left. \left. \left. ~~~~~~~~~~~~
-~ \frac{5}{6} \zeta(3) \alpha^2 + \frac{14591}{1296} \alpha^2
+ \zeta(3) \alpha^3 - \frac{1}{24} \alpha^3 - \frac{9163}{135} \zeta(3)
\right. \right. \right. \nonumber \\
&& \left. \left. \left. ~~~~~~~~~~~~
+~ 22 \zeta(4) - 20 \zeta(5) \right) C_F^2 C_A 
\right. \right. \nonumber \\
&& \left. \left. ~~~~~
+~ \left( \frac{265849}{7290} + \frac{410}{9} \zeta(3) \alpha 
- \frac{74771}{972} \alpha + \frac{26}{3} \zeta(3) \alpha^2 
\right. \right. \right. \nonumber \\
&& \left. \left. \left. ~~~~~~~~~~~~
-~ \frac{5063}{648} \alpha^2 - \frac{2}{3} \zeta(3) \alpha^3 
+ \frac{115}{216} \alpha^3 - \frac{15518}{405} \zeta(3)
\right. \right. \right. \nonumber \\
&& \left. \left. \left. ~~~~~~~~~~~~
-~ \frac{32}{3} \zeta(4) - 32 \zeta(5) \right) C_F^3 \right] a^3 \right] ~+~ 
O(a^4) 
\end{eqnarray} 
with further calculation giving 
\begin{eqnarray} 
\left. \Sigma^{(2) ~ {\mbox{\footnotesize{$\MSbar$}}} ~ 
\mbox{\footnotesize{finite}}}_{\bar{\psi} \sigma^{\mu\nu}  
D^\sigma D^\rho \psi}(p) \right|_{p^2 \, = \, \mu^2} &=& \frac{1}{2}  
\left. \Sigma^{(1) ~ {\mbox{\footnotesize{$\MSbar$}}} ~ 
\mbox{\footnotesize{finite}}}_{\bar{\psi} \sigma^{\mu\nu}  
D^\sigma D^\rho \psi}(p) \right|_{p^2 \, = \, \mu^2} ~+~ O(a^4) \nonumber \\ 
\left. \Sigma^{(3) ~ {\mbox{\footnotesize{$\MSbar$}}} ~ 
\mbox{\footnotesize{finite}}}_{\bar{\psi} \sigma^{\mu\nu}  
D^\sigma D^\rho \psi}(p) \right|_{p^2 \, = \, \mu^2} &=& -~ \frac{3}{2}  
\left. \Sigma^{(1) ~ {\mbox{\footnotesize{$\MSbar$}}} ~ 
\mbox{\footnotesize{finite}}}_{\bar{\psi} \sigma^{\mu\nu}  
D^\sigma D^\rho \psi}(p) \right|_{p^2 \, = \, \mu^2} ~+~ O(a^4) ~.  
\end{eqnarray} 
For the fourth moment of the transversity operator we have 
\begin{eqnarray} 
\left. \Sigma^{(1) ~ {\mbox{\footnotesize{$\MSbar$}}} ~ 
\mbox{\footnotesize{finite}}}_{\bar{\psi} \sigma^{\mu\nu}  
D^\sigma D^\rho D^\lambda \psi}(p) \right|_{p^2 \, = \, \mu^2} &=& 
-~ \frac{1}{32} ~+~ \left( \frac{293}{1152} + \frac{13}{384} \alpha \right) 
C_F a \nonumber \\
&& +~ \left[ \left( \frac{2037217}{829440} + \frac{3127}{18432} \alpha
+ \frac{397}{9216} \alpha^2 - \frac{13}{32} \zeta(3) \right) C_F C_A
\right. \nonumber \\
&& \left. ~~~~~
+~ \left( -~ \frac{8099}{5184} - \frac{1595}{4608} \alpha
+ \frac{29}{4608} \alpha^2 + \frac{1}{4} \zeta(3) \right) C_F^2
\right. \nonumber \\
&& \left. ~~~~~
-~ \frac{118621}{103680} \Nf T_F C_F
\right] a^2
\nonumber \\
&& +~ \left[ \left( -~ \frac{2440299949}{111974400}
+ \frac{59}{160} \zeta(3) \alpha - \frac{384991}{414720} \alpha
\right. \right. \nonumber \\
&& \left. \left. ~~~~~~
-~ \frac{14681}{6480} \zeta(3) - \frac{4}{3} \zeta(4) \right) \Nf T_F C_F C_A
\right. \nonumber \\
&& \left. ~~~~~
+~ \left( \frac{42119449}{11197440} - \frac{4}{9} \zeta(3) \alpha 
+ \frac{2369273}{1244160} \alpha
\right. \right. \nonumber \\
&& \left. \left. ~~~~~~~~~~~
+~ \frac{4291}{1080} \zeta(3) + \frac{4}{3} \zeta(4) \right) \Nf T_F C_F^2
\right. \nonumber \\
&& \left. ~~~~~
+~ \left( \frac{25433893}{6998400} + \frac{16}{81} \zeta(3) \right) 
\Nf^2 T_F^2 C_F
\right. \nonumber \\
&& \left. ~~~~~
+~ \left( \frac{112663077941}{3583180800} - \frac{99239}{69120} \zeta(3) \alpha 
- \frac{3}{256} \zeta(4) \alpha + \frac{5}{96} \zeta(5) \alpha 
\right. \right. \nonumber \\
&& \left. \left. ~~~~~~~~~~~
+~ \frac{89697707}{26542080} \alpha - \frac{2879}{27648} \zeta(3) \alpha^2 
- \frac{3}{512} \zeta(4) \alpha^2 
\right. \right. \nonumber \\
&& \left. \left. ~~~~~~~~~~~
-~ \frac{17}{384} \zeta(5) \alpha^2 + \frac{541433}{737280} \alpha^2
+ \frac{1}{96} \zeta(3) \alpha^3 + \frac{12979}{110592} \alpha^3
\right. \right. \nonumber \\
&& \left. \left. ~~~~~~~~~~~
-~ \frac{503123}{82944} \zeta(3) + \frac{181}{512} \zeta(4)
+ \frac{463}{384} \zeta(5) \right) C_F C_A^2 
\right. \nonumber \\
&& \left. ~~~~~
+~ \left( -~ \frac{2225988241}{89579520} + \frac{10093}{5760} \zeta(3) \alpha 
- \frac{16376411}{2488320} \alpha 
\right. \right. \nonumber \\
&& \left. \left. ~~~~~~~~~~~
+~ \frac{7}{96} \zeta(3) \alpha^2 - \frac{477515}{663552} \alpha^2 
- \frac{1}{32} \zeta(3) \alpha^3 - \frac{323}{12288} \alpha^3 
\right. \right. \nonumber \\
&& \left. \left. ~~~~~~~~~~~
+~ \frac{1219}{360} \zeta(3) - \frac{27}{32} \zeta(4) 
+ \frac{3}{4} \zeta(5) \right) C_F^2 C_A
\right. \nonumber \\
&& \left. ~~~~~
+~ \left( \frac{7904501}{3732480} - \frac{295}{144} \zeta(3) \alpha 
+ \frac{1045465}{248832} \alpha - \frac{1}{3} \zeta(3) \alpha^2 
\right. \right. \nonumber \\
&& \left. \left. ~~~~~~~~~~~
+~ \frac{64399}{165888} \alpha^2 + \frac{1}{48} \zeta(3) \alpha^3 
- \frac{1007}{55296} \alpha^3 + \frac{13739}{12960} \zeta(3)
\right. \right. \nonumber \\
&& \left. \left. ~~~~~~~~~~~
+~ \frac{7}{16} \zeta(4) + \frac{5}{6} \zeta(5) \right) C_F^3 
\right] a^3 ~+~ O(a^4) ~.  
\end{eqnarray} 
The remaining amplitudes are related to the first through 
\begin{eqnarray} 
\left. \Sigma^{(2) ~ {\mbox{\footnotesize{$\MSbar$}}} ~ 
\mbox{\footnotesize{finite}}}_{\bar{\psi} \sigma^{\mu\nu}  
D^\sigma D^\rho D^\lambda \psi}(p) \right|_{p^2 \, = \, \mu^2} &=& 
-~ \frac{4}{5} \left. \Sigma^{(1) ~ {\mbox{\footnotesize{$\MSbar$}}} ~ 
\mbox{\footnotesize{finite}}}_{\bar{\psi} \sigma^{\mu\nu}  
D^\sigma D^\rho D^\lambda \psi}(p) \right|_{p^2 \, = \, \mu^2} ~+~ O(a^4)  
\nonumber \\ 
\left. \Sigma^{(3) ~ {\mbox{\footnotesize{$\MSbar$}}} ~ 
\mbox{\footnotesize{finite}}}_{\bar{\psi} \sigma^{\mu\nu}  
D^\sigma D^\rho D^\lambda \psi}(p) \right|_{p^2 \, = \, \mu^2} &=& \frac{1}{5} 
\left. \Sigma^{(1) ~ {\mbox{\footnotesize{$\MSbar$}}} ~ 
\mbox{\footnotesize{finite}}}_{\bar{\psi} \sigma^{\mu\nu}  
D^\sigma D^\rho D^\lambda \psi}(p) \right|_{p^2 \, = \, \mu^2} ~+~ O(a^4)  
\end{eqnarray} 

Finally, for practical purposes we provide the general results for the specific
case of the Landau gauge when the colour group is $SU(3)$. Therefore, we have 
\begin{eqnarray} 
\left. \Sigma^{(1) ~ {\mbox{\footnotesize{$\MSbar$}}} ~ 
\mbox{\footnotesize{finite}}}_{\bar{\psi} \gamma^\mu 
D^\nu D^\sigma \psi}(p) \right|_{p^2 \, = \, \mu^2}^{SU(3), \alpha=0}
&=& \frac{4}{9} a ~+~ \left[ -~ \frac{50}{27} \Nf + \frac{4432}{135}
+ \frac{56}{15} \zeta(3) \right] a^2 \nonumber \\ 
&& +~ \left[ \frac{4210}{729} \Nf^2 + \left( -~ \frac{1665047}{7290}
- \frac{28}{5} \zeta(3) \right) \Nf \right. \nonumber \\
&& \left. ~~~~~ 
+~ \frac{279011797}{131220} - \frac{1717789}{2430} \zeta(3)
+ \frac{9370}{27} \zeta(5) \right] a^3 \nonumber \\ 
&& +~ O(a^4) 
\end{eqnarray} 
and 
\begin{eqnarray} 
\left. \Sigma^{(2) ~ {\mbox{\footnotesize{$\MSbar$}}} ~ 
\mbox{\footnotesize{finite}}}_{\bar{\psi} \gamma^\mu 
D^\nu D^\sigma \psi}(p) \right|_{p^2 \, = \, \mu^2}^{SU(3), \alpha=0}
&=& -~ \frac{1}{3} ~+~ \frac{214}{81} a ~+~ \left[ -~ \frac{32363}{5832} \Nf  
+ \frac{4763093}{87480} - \frac{152}{15} \zeta(3) \right] a^2  
\nonumber \\ 
&& +~ \left[ \left( \frac{1227463}{157464} + \frac{400}{729} \zeta(3) \right)
\Nf^2 \right. \nonumber \\
&& \left. ~~~~~ 
+~ \left( -~ \frac{1364405723}{4723920} - \frac{814}{405} \zeta(3)
- \frac{1000}{81} \zeta(4) \right) \Nf
\right. \nonumber \\
&& \left. ~~~~~ 
+~ \left( \frac{8619089351}{4723920} - \frac{12125507}{32805} \zeta(3)
\right. \right. \nonumber \\
&& \left. \left. ~~~~~~~~~~~
+~ \frac{8599}{972} \zeta(4) + \frac{2525}{27} \zeta(5) \right) \right] a^3 ~+~ 
O(a^4) ~.  
\end{eqnarray} 
For the RI$^\prime$ scheme we note that 
\begin{eqnarray} 
\left. \Sigma^{(1) ~ {\mbox{\footnotesize{RI$^\prime$}}} ~ 
\mbox{\footnotesize{finite}}}_{\bar{\psi} \gamma^\mu 
D^\nu D^\sigma \psi}(p) \right|_{p^2 \, = \, \mu^2}^{SU(3), \alpha=0}
&=& \frac{4}{9} a ~+~ \left[ -~ \frac{50}{27} \Nf + \frac{44168}{1215}
+ \frac{56}{15} \zeta(3) \right] a^2 \nonumber \\ 
&& +~ \left[ \frac{4210}{729} \Nf^2 + \left( -~ \frac{2738978}{10935}
- \frac{28}{5} \zeta(3) \right) \Nf \right. \nonumber \\
&& \left. ~~~~~ 
+~ \frac{326345791}{131220} - \frac{1678717}{2430} \zeta(3)
+ \frac{9370}{27} \zeta(5) \right] a^3 ~+~ O(a^4) \nonumber \\ 
\left. \Sigma^{(2) ~ {\mbox{\footnotesize{RI$^\prime$}}} ~ 
\mbox{\footnotesize{finite}}}_{\bar{\psi} \gamma^\mu 
D^\nu D^\sigma \psi}(p) \right|_{p^2 \, = \, \mu^2}^{SU(3), \alpha=0}
&=& -~ \frac{1}{3} ~+~ O(a^4) ~.  
\end{eqnarray} 
Further, for the next Wilson operator 
\begin{eqnarray} 
\left. \Sigma^{(1) ~ {\mbox{\footnotesize{$\MSbar$}}} ~ 
\mbox{\footnotesize{finite}}}_{\bar{\psi} \gamma^\mu 
D^\nu D^\sigma D^\rho \psi}(p) \right|_{p^2 \, = \, \mu^2}^{SU(3), \alpha=0}
&=& -~ 1 ~+~ \frac{7484}{675} a ~+~ \left[ -~ \frac{6041063}{243000} \Nf 
+ \frac{431713457}{1822500} - \frac{524}{15} \zeta(3) \right] a^2 \nonumber \\
&& +~ \left[ \left( \frac{1335574847}{32805000} + \frac{2512}{1215} \zeta(3)
\right) \Nf^2 \right. \nonumber \\
&& \left. ~~~~~
+~ \left( -~ \frac{1349388886469}{984150000} - \frac{43972}{1215} 
\zeta(3) - \frac{1256}{27} \zeta(4) \right) \Nf \right. \nonumber \\
&& \left. ~~~~~ 
+~ \frac{706189399771421}{78732000000} 
- \frac{112503104}{54675} \zeta(3) \right. \nonumber \\
&& \left. ~~~~~
+~ \frac{43507}{1620} \zeta(4) + \frac{7180}{9} \zeta(5) \right] a^3 ~+~ O(a^4) 
\end{eqnarray}  
and 
\begin{eqnarray} 
\left. \Sigma^{(2) ~ {\mbox{\footnotesize{$\MSbar$}}} ~ 
\mbox{\footnotesize{finite}}}_{\bar{\psi} \gamma^\mu 
D^\nu D^\sigma D^\rho \psi}(p) \right|_{p^2 \, = \, \mu^2}^{SU(3), \alpha=0}
&=& \frac{1}{40} a ~+~ \left[ -~ \frac{119}{1200} \Nf 
+ \frac{731129}{432000} + \frac{23}{90} \zeta(3) \right] a^2 \nonumber \\
&& +~ \left[ \frac{51959}{162000} \Nf^2 
+ \left( -~ \frac{232632277}{19440000} - \frac{755}{972} \zeta(3) \right) \Nf 
\right. \nonumber \\
&& \left. ~~~~~ 
+~ \frac{1047728166241}{9331200000} 
- \frac{109467991}{2916000} \zeta(3) + \frac{13111}{648} \zeta(5) \right] a^3
\nonumber \\ 
&& +~ O(a^4) ~.  
\end{eqnarray}  
As $\left. \Sigma^{(1) ~ {\mbox{\footnotesize{RI$^\prime$}}} ~ 
\mbox{\footnotesize{finite}}}_{\bar{\psi} \gamma^\mu 
D^\nu D^\sigma D^\rho \psi}(p) \right|_{p^2 \, = \, \mu^2}^{SU(3), 
\alpha=0}$~$=$~$(-1)$ by construction, we note 
\begin{eqnarray} 
\left. \Sigma^{(2) ~ {\mbox{\footnotesize{RI$^\prime$}}} ~ 
\mbox{\footnotesize{finite}}}_{\bar{\psi} \gamma^\mu 
D^\nu D^\sigma D^\rho \psi}(p) \right|_{p^2 \, = \, \mu^2}^{SU(3), \alpha=0}
&=& \frac{1}{40} a ~+~ \left[ -~ \frac{119}{1200} \Nf 
+ \frac{850873}{432000} + \frac{23}{90} \zeta(3) \right] a^2 \nonumber \\
&& +~ \left[ \frac{51959}{162000} \Nf^2 
+ \left( -~ \frac{266088707}{19440000} - \frac{755}{972} \zeta(3) \right) \Nf 
\right. \nonumber \\
&& \left. ~~~~~ 
+~ \frac{435587120587}{3110400000} 
- \frac{20750459}{583200} \zeta(3) + \frac{13111}{648} \zeta(5) \right] a^3
\nonumber \\ 
&& +~ O(a^4) ~.  
\end{eqnarray}  
For the transversity cases, when $n$~$=$~$3$ we have 
\begin{eqnarray} 
\left. \Sigma^{(1) ~ {\mbox{\footnotesize{$\MSbar$}}} ~ 
\mbox{\footnotesize{finite}}}_{\bar{\psi} \sigma^{\mu\nu} D^\sigma D^\rho 
\psi}(p) \right|_{p^2 \, = \, \mu^2}^{SU(3), \alpha=0}
&=& \frac{1}{18} \left[ 1 ~-~ \frac{218}{27} a ~+~ \left[ \frac{4394}{243} \Nf
- \frac{669202}{3645} + \frac{452}{15} \zeta(3) \right] a^2 \right.
\nonumber \\ 
&& \left. ~~~~~+~ \left[ \left( -~ \frac{177970}{6561} 
- \frac{416}{243} \zeta(3) \right) \Nf^2 
\right. \right. \nonumber \\
&& \left. \left. ~~~~~~~~~~~
+~ \left( \frac{98639141}{98415} + \frac{12712}{1215} \zeta(3) 
+ \frac{1040}{27} \zeta(4) \right) \Nf
\right. \right. \nonumber \\
&& \left. \left. ~~~~~~~~~~~
-~ \frac{1020141085}{157464} + \frac{59050063}{43740} \zeta(3)
\right. \right. \nonumber \\
&& \left. \left. ~~~~~~~~~~~
-~ \frac{7679}{324} \zeta(4) - \frac{12434}{27} \zeta(5) \right] a^3 
\right] ~+~ O(a^4)
\end{eqnarray} 
and 
\begin{equation}
\left. \Sigma^{(1) ~ {\mbox{\footnotesize{RI$^\prime$}}} ~ 
\mbox{\footnotesize{finite}}}_{\bar{\psi} \sigma^{\mu\nu} D^\sigma D^\rho
\psi}(p) \right|_{p^2 \, = \, \mu^2}^{SU(3), \alpha=0} ~=~ \frac{1}{18} ~+~
O(a^4) ~.
\end{equation} 
Finally, for the transversity moment $n$~$=$~$4$ we have 
\begin{eqnarray} 
\left. \Sigma^{(1) ~ {\mbox{\footnotesize{$\MSbar$}}} ~ 
\mbox{\footnotesize{finite}}}_{\bar{\psi} \sigma^{\mu\nu} D^\sigma D^\rho 
D^\lambda \psi}(p) \right|_{p^2 \, = \, \mu^2}^{SU(3), \alpha=0} &=& 
-~ \frac{1}{32} ~+~ \frac{293}{864} a ~+~ \left[ -~ \frac{118621}{155520} \Nf 
+ \frac{13151593}{1866240} - \frac{85}{72} \zeta(3) \right] a^2 
\nonumber \\
&& +~ \left[ \left( \frac{25433893}{20995200} + \frac{16}{243} \zeta(3) \right)
\Nf^2 \right. \nonumber \\
&& \left. ~~~~~+~ \left( -~ \frac{20277921581}{503884800}
- \frac{1943}{1944} \zeta(3) - \frac{40}{27} \zeta(4) \right) \Nf 
\right. \nonumber \\
&& \left. ~~~~~+~ \frac{2013899793847}{8062156800}
- \frac{146176079}{2799360} \zeta(3)
\right. \nonumber \\
&& \left. ~~~~~+~ \frac{2693}{3456} \zeta(4)
+ \frac{52991}{2592} \zeta(5) \right] a^3 ~+~ O(a^4) 
\end{eqnarray} 
with clearly
\begin{equation}
\left. \Sigma^{(1) ~ {\mbox{\footnotesize{RI$^\prime$}}} ~ 
\mbox{\footnotesize{finite}}}_{\bar{\psi} \sigma^{\mu\nu} D^\sigma D^\rho
D^\lambda \psi}(p) 
\right|_{p^2 \, = \, \mu^2}^{SU(3), \alpha=0} ~=~ -~ \frac{1}{32} ~+~ 
O(a^4) ~. 
\end{equation} 

\sect{Conversion functions.} 

An additional check on our computations is provided by the conversion functions
for each of the operators we have considered. These functions allow one to
convert the anomalous dimension of the operator in one renormalization scheme
to that in another scheme and are defined by the ratio of the renormalization
constants in both schemes 
\begin{equation} 
C_{\cal O}(a,\alpha) ~=~ 
\frac{Z^{\mbox{\footnotesize{RI$^\prime$}}}_{\cal O}} 
{Z^{\mbox{\footnotesize{$\MSbar$}}}_{\cal O}} ~. 
\end{equation} 
Then, \cite{37}, 
\begin{eqnarray}
\gamma^{\mbox{\footnotesize{RI$^\prime$}}}_{\cal O} 
\left(a_{\mbox{\footnotesize{RI$^\prime$}}}\right) &=& 
\gamma^{\mbox{\footnotesize{$\MSbar$}}}_{\cal O} 
\left(a_{\mbox{\footnotesize{$\MSbar$}}}\right) ~-~ 
\beta\left(a_{\mbox{\footnotesize{$\MSbar$}}}\right) 
\frac{\partial ~}{\partial a_{\mbox{\footnotesize{$\MSbar$}}}} 
\ln C_{\cal O} \left(a_{\mbox{\footnotesize{$\MSbar$}}}, 
\alpha_{\mbox{\footnotesize{$\MSbar$}}}\right) \nonumber \\
&& -~ \alpha_{\mbox{\footnotesize{$\MSbar$}}} 
\gamma^{\mbox{\footnotesize{$\MSbar$}}}_\alpha 
\left(a_{\mbox{\footnotesize{$\MSbar$}}}\right) 
\frac{\partial ~}{\partial \alpha_{\mbox{\footnotesize{$\MSbar$}}}}  
\ln C_{\cal O} \left(a_{\mbox{\footnotesize{$\MSbar$}}},  
\alpha_{\mbox{\footnotesize{$\MSbar$}}}\right) 
\end{eqnarray} 
where one needs to express the $\MSbar$ variables in terms of the RI$^\prime$
scheme using (\ref{cccon}) and (\ref{alpcon}) in order to compare with the 
anomalous dimensions from the explicit computation. We record that the 
conversion functions for the various operators we are interested in here are 
\begin{eqnarray} 
C_{\bar{\psi} \gamma^\mu D^\nu D^\sigma \psi}(a,\alpha) &=& 1 ~+~ \left( 27 
\alpha + 107 \right) \frac{C_F a}{18} \nonumber \\
&& +~ \left[ \left( 86400 \alpha^2 - 93312 \zeta(3) \alpha + 409104 \alpha 
- 715392 \zeta(3) + 3302464 \right) C_A \right. \nonumber \\
&& \left. ~~~~+~ \left( 60480 \alpha^2 + 327600 \alpha + 373248 \zeta(3) 
+ 327549 \right) C_F \right. \nonumber \\  
&& \left. ~~~~-~ 1475960 T_F \Nf \right] \frac{C_F a^2}{51840} \nonumber \\
&& +~ \left[ \left( 11032200 \alpha^3 - 13915152 \zeta(3) \alpha^2  
- 466560 \zeta(5) \alpha^2 + 64538856 \alpha^2 \right. \right. \nonumber \\
&& \left. \left. ~~~~~~~-~ 141379344 \zeta(3) \alpha + 8398080 \zeta(5) \alpha
+ 319887792 \alpha \right. \right. \nonumber \\
&& \left. \left. ~~~~~~~-~ 635381280 \zeta(3) + 25660800 \zeta(4) 
+ 142767360 \zeta(5) \right. \right. \nonumber \\
&& \left. \left. ~~~~~~~+~ 2356357048 \right) C_A^2 \right. \nonumber \\
&& \left. ~~~~+~ \left( 4607280 \alpha^3 + 5785344 \zeta(3) \alpha^2 
+ 32256792 \alpha^2 - 13841280 \zeta(3) \alpha \right. \right. \nonumber \\
&& \left. \left. ~~~~~~~~~~+~ 33592320 \zeta(5) \alpha + 180233856 \alpha 
- 297743040 \zeta(3) \right. \right. \nonumber \\
&& \left. \left. ~~~~~~~~~~-~ 76982400 \zeta(4) - 59719680 \zeta(5) 
+ 1051633031 \right) C_A C_F \right. \nonumber \\
&& \left. ~~~~+~ \left( 35085312 \zeta(3) \alpha - 97555104 \alpha 
- 75098880 \zeta(3) \right. \right. \nonumber \\
&& \left. \left. ~~~~~~~~~~-~ 93312000 \zeta(4) - 1625432200 \right) C_A T_F 
\Nf \right. \nonumber \\
&& \left. ~~~~+~ \left( 2177280 \alpha^3 - 23328000 \zeta(3) \alpha^2 
+ 27799200 \alpha^2 - 73685376 \zeta(3) \alpha \right. \right. \nonumber \\
&& \left. \left. ~~~~~~~~~~+~ 90339057 \alpha + 319139136 \zeta(3) 
+ 51321600 \zeta(4) \right. \right. \nonumber \\
&& \left. \left. ~~~~~~~~~~+~ 201553920 \zeta(5) - 607345686 \right) C_F^2 
\right. \nonumber \\
&& \left. ~~~~-~ \left( 31104000 \zeta(3) \alpha + 63327960 \alpha  
- 303948288 \zeta(3) \right. \right. \nonumber \\
&& \left. \left. ~~~~~~~~~~-~ 93312000 \zeta(4) + 922104436 \right) C_F T_F \Nf 
\right. \nonumber \\
&& \left. ~~~~+~ \left( 13824000 \zeta(3) + 250653280 \right) T_F^2 \Nf^2 
\right] \frac{C_F a^3}{2799360} ~+~ O(a^4)  
\end{eqnarray} 
\begin{eqnarray} 
C_{\bar{\psi} \gamma^\mu D^\nu D^\sigma D^\rho \psi}(a,\alpha) &=& 1 ~+~ 
\left( 525 \alpha + 1871 \right) \frac{C_F a}{225} \nonumber \\
&& +~ \left[ \left( 18315000 \alpha^2 - 23328000 \zeta(3) \alpha + 88528500
\alpha - 103680000 \zeta(3) \right. \right. \nonumber \\
&& \left. \left. ~~~~~~+~ 603802180 \right) C_A \right. \nonumber \\
&& \left. ~~~~+~ \left( 21330000 \alpha^2 + 119182200 \alpha 
+ 62208000 \zeta(3) + 98349057 \right) C_F \right. \nonumber \\  
&& \left. ~~~~-~ 264322520 T_F \Nf \right] \frac{C_F a^2}{6480000} \nonumber \\
&& +~ \left[ \left( 239334750000 \alpha^3 - 340977600000 \zeta(3) \alpha^2  
- 2916000000 \zeta(5) \alpha^2 \right. \right. \nonumber \\
&& \left. \left. ~~~~~~+~ 1428857212500 \alpha^2 
- 3356364600000 \zeta(3) \alpha \right. \right. \nonumber \\
&& \left. \left. ~~~~~~+~ 174960000000 \zeta(5) \alpha 
+ 7142849746875 \alpha \right. \right. \nonumber \\
&& \left. \left. ~~~~~~-~ 12700608624000 \zeta(3) + 335689920000 \zeta(4) 
\right. \right. \nonumber \\
&& \left. \left. ~~~~~~+~ 2504844000000 \zeta(5) 
+ 48069511158775 \right) C_A^2 \right. \nonumber \\
&& \left. ~~~~+~ \left( 229047750000 \alpha^3 - 142300800000 \zeta(3) \alpha^2 
+ 1689792435000 \alpha^2 \right. \right. \nonumber \\
&& \left. \left. ~~~~~~~~~~-~ 1524733632000 \zeta(3) \alpha 
+ 839808000000 \zeta(5) \alpha \right. \right. \nonumber \\
&& \left. \left. ~~~~~~~~~~+~ 9165889557000 \alpha - 1770530400000 \zeta(3) 
\right. \right. \nonumber \\
&& \left. \left. ~~~~~~~~~~-~ 1007069760000 \zeta(4) + 653184000000 \zeta(5) 
\right. \right. \nonumber \\
&& \left. \left. ~~~~~~~~~~+~ 18744964493980 \right) C_A C_F \right. 
\nonumber \\
&& \left. ~~~~+~ \left( 816480000000 \zeta(3) \alpha - 2158137000000 \alpha 
- 1801163520000 \zeta(3) \right. \right. \nonumber \\
&& \left. \left. ~~~~~~~~~~-~ 1464998400000 \zeta(4) - 31372620527200 \right) 
C_A T_F \Nf \right. \nonumber \\
&& \left. ~~~~+~ \left( 145435500000 \alpha^3 - 366249600000 \zeta(3) \alpha^2 
+ 1372611420000 \alpha^2 \right. \right. \nonumber \\
&& \left. \left. ~~~~~~~~~~-~ 1048904640000 \zeta(3) \alpha 
+ 3148063990200 \alpha \right. \right. \nonumber \\
&& \left. \left. ~~~~~~~~~~+~ 4800009888000 \zeta(3) 
+ 671379840000 \zeta(4) \right. \right. \nonumber \\
&& \left. \left. ~~~~~~~~~~+~ 3359232000000 \zeta(5) - 8872064364708 \right) 
C_F^2 \right. \nonumber \\
&& \left. ~~~~-~ \left( 488332800000 \zeta(3) \alpha + 2684380392000 \alpha  
- 4552485120000 \zeta(3) \right. \right. \nonumber \\
&& \left. \left. ~~~~~~~~~~-~ 1464998400000 \zeta(4) + 18121905428720 \right) 
C_F T_F \Nf \right. \nonumber \\
&& \left. ~~~~+~ \left( 217036800000 \zeta(3) + 4952079510400 \right) T_F^2 
\Nf^2 \right] \frac{C_F a^3}{34992000000} \nonumber \\
&& ~+~ O(a^4)  
\end{eqnarray} 
\begin{eqnarray} 
C_{\bar{\psi} \sigma^{\mu\nu} D^\sigma D^\rho \psi}(a,\alpha) &=& 1 ~+~ \left( 
33 \alpha + 109 \right) \frac{C_F a}{18} \nonumber \\
&& +~ \left[ \left( 6600 \alpha^2 - 7776 \zeta(3) \alpha + 32067 \alpha 
- 47952 \zeta(3) + 228974 \right) C_A \right. \nonumber \\
&& \left. ~~~~+~ \left( 6480 \alpha^2 + 30900 \alpha + 31104 \zeta(3) 
+ 10917 \right) C_F \right. \nonumber \\  
&& \left. ~~~~-~ 99220 T_F \Nf \right] \frac{C_F a^2}{3240} \nonumber \\
&& +~ \left[ \left( 3407670 \alpha^3 - 4575204 \zeta(3) \alpha^2  
- 58320 \zeta(5) \alpha^2 + 20121777 \alpha^2 \right. \right. \nonumber \\
&& \left. \left. ~~~~~~~-~ 46022256 \zeta(3) \alpha + 2799360 \zeta(5) \alpha
+ 100170405 \alpha \right. \right. \nonumber \\
&& \left. \left. ~~~~~~~-~ 196675020 \zeta(3) + 3732480 \zeta(4) 
+ 45081360 \zeta(5) \right. \right. \nonumber \\
&& \left. \left. ~~~~~~~+~ 693358478 \right) C_A^2 \right. \nonumber \\
&& \left. ~~~~+~ \left( 2334420 \alpha^3 - 46656 \zeta(3) \alpha^2 
+ 15956352 \alpha^2 - 12324960 \zeta(3) \alpha \right. \right. \nonumber \\
&& \left. \left. ~~~~~~~~~~+~ 11197440 \zeta(5) \alpha + 86296752 \alpha 
- 34434720 \zeta(3) \right. \right. \nonumber \\
&& \left. \left. ~~~~~~~~~~-~ 11197440 \zeta(4) 
+ 214883180 \right) C_A C_F \right. \nonumber \\
&& \left. ~~~~+~ \left( 11384064 \zeta(3) \alpha - 30726648 \alpha 
- 17280000 \zeta(3) \right. \right. \nonumber \\
&& \left. \left. ~~~~~~~~~~-~ 24261120 \zeta(4) - 463372640 \right) C_A T_F 
\Nf \right. \nonumber \\
&& \left. ~~~~+~ \left( 1399680 \alpha^3 - 6065280 \zeta(3) \alpha^2 
+ 13024800 \alpha^2 - 13965696 \zeta(3) \alpha \right. \right. \nonumber \\
&& \left. \left. ~~~~~~~~~~+~ 26888652 \alpha + 108183168 \zeta(3) 
+ 7464960 \zeta(4) \right. \right. \nonumber \\
&& \left. \left. ~~~~~~~~~~+~ 22394880 \zeta(5) - 153974772 \right) C_F^2 
\right. \nonumber \\
&& \left. ~~~~-~ \left( 8087040 \zeta(3) \alpha + 27287280 \alpha  
- 69133824 \zeta(3) \right. \right. \nonumber \\
&& \left. \left. ~~~~~~~~~~-~ 24261120 \zeta(4) + 231549328 \right) C_F T_F \Nf 
\right. \nonumber \\
&& \left. ~~~~+~ \left( 3594240 \zeta(3) + 70515200 \right) T_F^2 \Nf^2 
\right] \frac{C_F a^3}{699840} ~+~ O(a^4)  
\end{eqnarray} 
and
\begin{eqnarray} 
C_{\bar{\psi} \sigma^{\mu\nu} D^\sigma D^\rho D^\lambda \psi}(a,\alpha) &=& 
1 ~+~ \left( 75 \alpha + 293 \right) \frac{C_F a}{36} \nonumber \\
&& +~ \left[ \left( 64890 \alpha^2 - 77760 \zeta(3) \alpha + 309195 \alpha 
- 414720 \zeta(3) + 2302897 \right) C_A \right. \nonumber \\
&& \left. ~~~~+~ \left( 63720 \alpha^2 + 380940 \alpha + 207360 \zeta(3) 
+ 404940 \right) C_F \right. \nonumber \\  
&& \left. ~~~~-~ 1039688 T_F \Nf \right] \frac{C_F a^2}{25920} \nonumber \\
&& +~ \left[ \left( 677127600 \alpha^3 - 918993600 \zeta(3) \alpha^2  
- 18662400 \zeta(5) \alpha^2 \right. \right. \nonumber \\
&& \left. \left. ~~~~~~~+~ 4008299580 \alpha^2 - 9063653760 \zeta(3) \alpha 
+ 466560000 \zeta(5) \alpha \right. \right. \nonumber \\
&& \left. \left. ~~~~~~~+~ 19846116045 \alpha - 36380232000 \zeta(3) 
+ 783820800 \zeta(4) \right. \right. \nonumber \\
&& \left. \left. ~~~~~~~+~ 8939289600 \zeta(5) 
+ 140182687541 \right) C_A^2 \right. \nonumber \\
&& \left. ~~~~+~ \left( 517071600 \alpha^3 - 102643200 \zeta(3) \alpha^2 
+ 3840647400 \alpha^2 \right. \right. \nonumber \\
&& \left. \left. ~~~~~~~~~~-~ 3332482560 \zeta(3) \alpha 
+ 2239488000 \zeta(5) \alpha + 21797212320 \alpha \right. \right. \nonumber \\
&& \left. \left. ~~~~~~~~~~-~ 9369146880 \zeta(3) - 2351462400 \zeta(4) 
\right. \right. \nonumber \\
&& \left. \left. ~~~~~~~~~~+~ 447897600 \zeta(5)
+ 58907275400 \right) C_A C_F \right. \nonumber \\
&& \left. ~~~~+~ \left( 2217093120 \zeta(3) \alpha - 6005931840 \alpha 
- 6177116160 \zeta(3) \right. \right. \nonumber \\
&& \left. \left. ~~~~~~~~~~-~ 4777574400 \zeta(4) - 94522187168 \right) C_A T_F 
\Nf \right. \nonumber \\
&& \left. ~~~~+~ \left( 279936000 \alpha^3 - 1194393600 \zeta(3) \alpha^2 
+ 3013848000 \alpha^2 \right. \right. \nonumber \\
&& \left. \left. ~~~~~~~~~~-~ 4503859200 \zeta(3) \alpha + 8684647200 \alpha 
+ 18380113920 \zeta(3) \right. \right. \nonumber \\
&& \left. \left. ~~~~~~~~~~+~ 1567641600 \zeta(4) + 2985984000 \zeta(5) 
- 24416900640 \right) C_F^2 \right. \nonumber \\
&& \left. ~~~~-~ \left( 1592524800 \zeta(3) \alpha + 6750907200 \alpha  
- 16028098560 \zeta(3) \right. \right. \nonumber \\
&& \left. \left. ~~~~~~~~~~-~ 4777574400 \zeta(4) + 57917250880 \right) 
C_F T_F \Nf \right. \nonumber \\
&& \left. ~~~~+~ \left( 707788800 \zeta(3) + 15192521216 \right) T_F^2 \Nf^2 
\right] \frac{C_F a^3}{111974400} \nonumber \\
&& +~ O(a^4) 
\end{eqnarray} 
where the coupling constant and gauge parameter are in the $\MSbar$ scheme. We 
note that the same RI$^\prime$ anomalous dimensions are determined as
previously. 

\sect{Discussion.} 
We have provided the finite parts of various Green's functions required for
the renormalization of the $n$~$=$~$3$ and $4$ moments of the non-singlet
twist-$2$ Wilson and transversity operators at three loops in both the 
$\MSbar$ and RI$^\prime$ schemes. Since these are available at several loop
orders, the hope is that they will be central to the extraction of accurate
values for the matrix elements which will be measured on the lattice. From 
another point of view the new $\MSbar$ anomalous dimensions which are now 
available for the moments up to and including $4$ for the transversity operator
will provide a useful check on the full $n$-dependent three loop transversity 
anomalous dimensions when they are eventually determined. The impressive 
symbolic manipulation machinery which achieved the arbitrary $n$ anomalous 
dimensions for the twist-$2$ flavour non-singlet and singlet Wilson operators, 
\cite{1,2,3,4}, can be applied to the transversity case. Whilst this can be 
achieved in principle, in the interim one could follow a similar direction to 
the earlier approach of \cite{39,40} where the anomalous dimensions of the 
Wilson operators were determined to moment $n$~$=$~$10$ and later to higher 
moments, $n$~$\leq$~$16$ (except $n$~$=$~$14$), \cite{45,46}. These explicit 
moments were then used to construct solid approximations to the full anomalous 
dimensions which were then shown to be credible in a substantial range of the 
$x$ variable in, for example, \cite{47}. Given the advance in computer 
capabilities since \cite{39,40}, it would seem to us that a fixed moment 
computation for the anomalous dimensions of the higher moments of the
transversity operator is certainly viable. Moreover, one would not be 
constrained by the choice of an arbitrary covariant gauge and the use of the 
Feynman gauge would therefore reduce computer time.  

\vspace{1cm}
\noindent 
{\bf Acknowledgements.} The author thanks Dr R. Horsley, Dr P.E.L. Rakow and Dr
C. McNeile for valuable discussions. 

\appendix

\sect{Projection tensors.}
In this appendix we record the decomposition of the various Green's functions
we have computed into the various tensor bases. For the Wilson operators there
are two independent tensors consistent with the symmetries and tracelessness
conditions of the original operator. By contrast the Green's functions
involving the transversity operator require three independent tensors. First,
for the Wilson operators, for $n$~$=$~$3$, we define  
\begin{eqnarray}
\langle \psi(-p) {\cal O}_W^{\mu\nu\sigma}(0) \bar{\psi}(p) \rangle &=& 
\Sigma^{(1)}_{\bar{\psi} \gamma^\mu D^\nu D^\sigma \psi} (p) 
\left[ p^\mu p^\nu p^\sigma \pslash 
- \frac{p^2}{(d+2)} \eta^{(\mu\nu} p^{\sigma)} \pslash \right] 
\frac{1}{(p^2)^3} \nonumber \\
&& +~ \Sigma^{(2)}_{\bar{\psi} \gamma^\mu D^\nu D^\sigma \psi} (p) \! 
\left[ \gamma^{(\mu} p^\nu p^{\sigma)} - \frac{2}{(d+2)} \eta^{(\mu\nu} 
p^{\sigma)} \pslash \right. \nonumber \\
&& \left. ~~~~~~~~~~~~~~~~~~~~~~~-~ \frac{p^2}{(d+2)} \eta^{(\mu\nu} 
\gamma^{\sigma)} \right] \frac{1}{(p^2)^3} ~.  
\end{eqnarray}
Then 
\begin{eqnarray}
\Sigma^{(1)}_{\bar{\psi} \gamma^\mu D^\nu D^\sigma \psi} (p) &=& \left[
\frac{(d+2)(d+4)}{4(d^2-1)} \left[ p^\mu p^\nu p^\sigma \pslash
- \frac{p^2}{(d+2)} \eta^{(\mu\nu} p^{\sigma)} \pslash \right] \right.
\nonumber \\
&& \left. ~-~ \frac{(d+2)p^2}{4(d^2-1)} \left[ \gamma^{(\mu} p^\nu p^{\sigma)}
- \frac{2}{(d+2)} \eta^{(\mu\nu} p^{\sigma)} \pslash 
- \frac{p^2}{(d+2)} \eta^{(\mu\nu} \gamma^{\sigma)} \right] \right] 
\nonumber \\  
&& \times ~ \frac{1}{(p^2)^3} 
\langle \psi(-p) {\cal O}_W^{\mu\nu\sigma}(0) \bar{\psi}(p) \rangle 
\end{eqnarray}
and 
\begin{eqnarray}
\Sigma^{(2)}_{\bar{\psi} \gamma^\mu D^\nu D^\sigma \psi} (p) &=& \left[
-~ \frac{(d+2)}{4(d^2-1)} \left[ p^\mu p^\nu p^\sigma \pslash
- \frac{p^2}{(d+2)} \eta^{(\mu\nu} p^{\sigma)} \pslash \right] \right.
\nonumber \\
&& \left. ~+~ \frac{(d+2)p^2}{12(d^2-1)} \left[ \gamma^{(\mu} p^\nu p^{\sigma)}
- \frac{2}{(d+2)} \eta^{(\mu\nu} p^{\sigma)} \pslash 
- \frac{p^2}{(d+2)} \eta^{(\mu\nu} \gamma^{\sigma)} \right] \right] 
\nonumber \\ 
&& \times ~ \frac{1}{(p^2)^3} 
\langle \psi(-p) {\cal O}_W^{\mu\nu\sigma}(0) \bar{\psi}(p) \rangle ~.  
\end{eqnarray}
In this appendix and the next in order to compactify our expressions, we note 
that the symmetrization on Lorentz indices using parentheses excludes the 
standard normalization factor. So, for instance,
\begin{equation}
X^{(\mu\nu\sigma)} ~=~ X^{\mu\nu\sigma} ~+~ X^{\nu\sigma\mu} ~+~
X^{\sigma\mu\nu} ~+~ X^{\mu\sigma\nu} ~+~ X^{\sigma\nu\mu} ~+~ 
X^{\nu\mu\sigma} ~. 
\end{equation}
Further, in our convention when this operation involves a tensor which is
itself symmetric, then one only counts the independent object once. So, for
example,
\begin{equation}
\eta^{\mu(\nu} \eta^{\sigma\rho)} ~=~  \eta^{(\mu\nu} \eta^{\sigma\rho)} ~=~  
\eta^{\mu\nu} \eta^{\sigma\rho} ~+~ \eta^{\mu\sigma} \eta^{\nu\rho} ~+~ 
\eta^{\mu\rho} \eta^{\nu\sigma} ~.
\end{equation} 
For the $n$~$=$~$4$ moment of the Wilson operator we decompose the Green's
function into two independent tensors with 
\begin{equation}
\langle \psi(-p) {\cal O}_W^{\mu\nu\sigma\rho}(0) \bar{\psi}(p) \rangle ~=~ 
\sum_{i=1}^2 \bar{\Sigma}^{(i)}_{\bar{\psi} \gamma^\mu D^\nu D^\sigma D^\rho 
\psi}(p) W^{\mu\nu\sigma\rho}_{i}(p) ~. 
\end{equation} 
Then each individual amplitude is given by the projections 
\begin{eqnarray} 
\bar{\Sigma}^{(1)}_{\bar{\psi} \gamma^\mu D^\nu D^\sigma D^\rho \psi} (p) &=& 
\left[ \frac{1}{4(d+4)(d+2)(d^2-1)} W^{\mu\nu\sigma\rho}_{1}(p) ~-~ 
\frac{1}{4(d+4)(d^2-1)p^2} W^{\mu\nu\sigma\rho}_{2}(p) \right] \nonumber \\
&& \times ~ \langle \psi(-p) {\cal O}_{W\,\mu\nu\sigma\rho}(0) \bar{\psi}(p) 
\rangle \nonumber \\  
\bar{\Sigma}^{(2)}_{\bar{\psi} \gamma^\mu D^\nu D^\sigma D^\rho \psi} (p) &=& 
\left[ -~ \frac{1}{4(d+4)(d^2-1)} W^{\mu\nu\sigma\rho}_{1}(p) ~+~ 
\frac{(d+3)}{4(d+4)(d^2-1)p^2} W^{\mu\nu\sigma\rho}_{2}(p) \right] 
\nonumber \\
&& \times ~ \langle \psi(-p) {\cal O}_{W\,\mu\nu\sigma\rho}(0) \bar{\psi}(p) 
\rangle 
\end{eqnarray} 
where the two basis tensors are 
\begin{eqnarray}
W^{\mu\nu\sigma\rho}_1(p) &=& \eta^{(\mu\nu} \eta^{\sigma\rho)} \pslash ~-~
\frac{(d+2)}{2} \eta^{(\mu\nu} p^\sigma \gamma^{\rho)} ~+~ 
\frac{(d+2)(d+4)}{2p^2} p^{(\mu} p^\nu p^\sigma \gamma^{\rho)} \nonumber \\
&& -~ \frac{(d+2)(d+4)}{(p^2)^2} p^\mu p^\nu p^\sigma p^\rho \pslash 
\nonumber \\  
W^{\mu\nu\sigma\rho}_2(p) &=& \eta^{(\mu\nu} p^\sigma p^{\rho)} \pslash ~-~
\frac{p^2}{2} \eta^{(\mu\nu} p^\sigma \gamma^{\rho)} ~+~ \frac{(d+4)}{2}
p^{(\mu} p^\nu p^\sigma \gamma^{\rho)} \nonumber \\
&& -~ \frac{2(d+4)}{p^2} p^\mu p^\nu p^\sigma p^\rho \pslash ~. 
\end{eqnarray}  
Finally, for this particular case to ensure that the tree term of one of the 
amplitudes involves unity, we have chosen to redefine the amplitudes as 
\begin{eqnarray}
\bar{\Sigma}^{(1)}_{\bar{\psi} \gamma^\mu D^\nu D^\sigma D^\rho \psi} (p) &=& 
-~ (d+2)(d+4) \left[ \Sigma^{(1)}_{\bar{\psi} \gamma^\mu D^\nu D^\sigma D^\rho 
\psi} (p) ~-~ \frac{1}{2(d+4)} \Sigma^{(2)}_{\bar{\psi} \gamma^\mu D^\nu 
D^\sigma D^\rho \psi} (p) \right] \nonumber \\ 
\bar{\Sigma}^{(2)}_{\bar{\psi} \gamma^\mu D^\nu D^\sigma D^\rho \psi} (p) &=& 
\Sigma^{(2)}_{\bar{\psi} \gamma^\mu D^\nu D^\sigma D^\rho \psi} (p) ~. 
\end{eqnarray}

Turning to the cases for the transversity operators then for $n$~$=$~$3$ the 
decomposition is 
\begin{equation}
\langle \psi(-p) {\cal O}_T^{\mu\nu\sigma\rho}(0) \bar{\psi}(p) \rangle ~=~ 
\sum_{i=1}^3 \Sigma^{(i)}_{\bar{\psi} \sigma^{\mu\nu} D^\sigma D^\rho \psi} (p)
T^{(3)\mu\nu\sigma\rho}_{i}(p) ~.
\end{equation} 
Hence the coefficients are given by the three projections of the Green's 
function 
\begin{eqnarray} 
\Sigma^{(1)}_{\bar{\psi} \sigma^{\mu\nu} D^\sigma D^\rho \psi} (p) &=& 
\left[ -~ \frac{1}{12(d^2-4)(d^2-1)} T^{(3)\mu\nu\sigma\rho}_{1}(p) ~-~ 
\frac{1}{12(d^2-4)(d^2-1)} T^{(3)\mu\nu\sigma\rho}_{2}(p) \right] \nonumber \\ 
&& \times ~ \langle \psi(-p) {\cal O}_{T\,\mu\nu\sigma\rho}(0) \bar{\psi}(p) 
\rangle \nonumber \\  
\Sigma^{(2)}_{\bar{\psi} \sigma^{\mu\nu} D^\sigma D^\rho \psi} (p) &=& 
\left[ -~ \frac{1}{12(d^2-4)(d^2-1)} T^{(3)\mu\nu\sigma\rho}_{1}(p) \right.
\nonumber \\
&& \left. ~-~ \frac{(d^3+d^2-2d+4)}{12d(d+4)(d^2-4)(d^2-1)} 
T^{(3)\mu\nu\sigma\rho}_{2}(p) \right. \nonumber \\ 
&& \left. ~-~ \frac{1}{12d(d+4)(d-1)} T^{(3)\mu\nu\sigma\rho}_{3}(p) \right] 
\langle \psi(-p) {\cal O}_{T\,\mu\nu\sigma\rho}(0) \bar{\psi}(p) \rangle 
\nonumber \\ 
\Sigma^{(3)}_{\bar{\psi} \sigma^{\mu\nu} D^\sigma D^\rho \psi} (p) &=& 
\left[ -\, \frac{1}{12d(d+4)(d-1)} T^{(3)\mu\nu\sigma\rho}_{2}(p) 
\,-\, \frac{(d^2+2)}{12d(d+4)(d-2)(d^2-1)} T^{(3)\mu\nu\sigma\rho}_{3}(p) 
\right] \nonumber \\
&& \times ~ \langle \psi(-p) {\cal O}_{T\,\mu\nu\sigma\rho}(0) \bar{\psi}(p) 
\rangle 
\end{eqnarray} 
where the choice of basis tensors is 
\begin{eqnarray} 
T^{(3)\mu\nu\sigma\rho}_{1}(p) &=& 
-~ \frac{(d+2)(d+4)}{p^2} \sigma^{\lambda\mu} p^\nu p^\sigma p^\rho 
p_\lambda ~+~ (d+2) \sigma^{\lambda\mu} \eta^{(\nu\sigma} p^{\rho)} p_\lambda  
\nonumber \\
&& -~ (d+2) \sigma^{\mu(\nu} p^\sigma p^{\rho)} ~+~ \sigma^{\mu(\nu} 
\eta^{\sigma\rho)} p^2 \nonumber \\  
T^{(3)\mu\nu\sigma\rho}_{2}(p) &=& 
\frac{(d+1)(d+2)}{p^2} \sigma^{\lambda\mu} p^\nu p^\sigma p^\rho 
p_\lambda ~-~ (d+1) \sigma^{\lambda\mu} \eta^{(\nu\sigma} p^{\rho)} p_\lambda  
\nonumber \\
&& +~ \frac{(d+2)}{p^2} \sigma^{\lambda(\nu} p^\sigma p^{\rho)} p^\mu 
p_\lambda  ~-~ \sigma^{\mu(\nu} \eta^{\sigma\rho)} p^2 \nonumber \\ 
T^{(3)\mu\nu\sigma\rho}_{3}(p) &=& 
-~ \frac{(d^2+2d-2)}{p^2} \sigma^{\lambda\mu} p^\nu p^\sigma p^\rho 
p_\lambda ~+~ d \sigma^{\lambda\mu} \eta^{(\nu\sigma} p^{\rho)} p_\lambda  
\nonumber \\
&& +~ \frac{2}{p^2} \sigma^{\lambda(\nu} p^\sigma p^{\rho)} p^\mu 
p_\lambda  ~-~ \sigma^{\lambda(\nu} \eta^{|\mu|\sigma} p^{\rho)} p_\lambda ~.  
\end{eqnarray} 
Finally, for the $n$~$=$~$4$ moment of the transversity operator we set 
\begin{equation}
\langle \psi(-p) {\cal O}_T^{\mu\nu\sigma\rho\lambda}(0) \bar{\psi}(p) 
\rangle ~=~ \sum_{i=1}^3 \Sigma^{(i)}_{\bar{\psi} \sigma^{\mu\nu} D^\sigma 
D^\rho D^\lambda \psi} (p) T^{\mu\nu\sigma\rho\lambda}_{i}(p)
\end{equation} 
where the three amplitudes are given by 
\begin{eqnarray} 
\Sigma^{(1)}_{\bar{\psi} \sigma^{\mu\nu} D^\sigma D^\rho D^\lambda \psi} (p) 
&=& -~ \frac{1}{16(d+4)(d^2-1)(d-2)} T^{\mu\nu\sigma\rho\lambda}_{1}(p)
\langle \psi(-p) {\cal O}_{T\,\mu\nu\sigma\rho\lambda}(0) \bar{\psi}(p) \rangle 
\nonumber \\  
\Sigma^{(2)}_{\bar{\psi} \sigma^{\mu\nu} D^\sigma D^\rho D^\lambda \psi} (p) 
&=& \left[ -~ \frac{(d^2+7d+14)}{4(d+6)(d+4)(d+3)(d+1)(d^2-1)(d-2)} 
T^{\mu\nu\sigma\rho\lambda}_{2}(p) \right. \nonumber \\
&& \left. ~+~ \frac{(d+2)}{2(d+6)(d+4)(d+3)(d+1)(d^2-1)(d-2)} 
T^{\mu\nu\sigma\rho\lambda}_{3}(p) \right] \nonumber \\
&& \times ~ \langle \psi(-p) {\cal O}_{T\,\mu\nu\sigma\rho\lambda}(0) 
\bar{\psi}(p) \rangle \nonumber \\  
\Sigma^{(3)}_{\bar{\psi} \sigma^{\mu\nu} D^\sigma D^\rho D^\lambda \psi} (p) 
&=& \left[ \frac{(d+2)}{2(d+6)(d+4)(d+3)(d+1)(d^2-1)(d-2)} 
T^{\mu\nu\sigma\rho\lambda}_{2}(p) \right. \nonumber \\
&& \left. ~-~ \frac{(d^2+d+2)}{16(d+6)(d+4)(d+3)(d+1)(d^2-1)(d-2)} 
T^{\mu\nu\sigma\rho\lambda}_{3}(p) \right] \nonumber \\
&& \times ~ \langle \psi(-p) {\cal O}_{T\,\mu\nu\sigma\rho\lambda}(0) 
\bar{\psi}(p) \rangle 
\end{eqnarray} 
in terms of the three basis tensors 
\begin{eqnarray} 
T^{\mu\nu\sigma\rho\lambda}_{1}(p) &=& 
-~ \frac{4(d+4)}{p^2} \sigma^{\theta\mu} p^\nu p^\sigma p^\rho p^\lambda 
p_\theta ~+~ 2 \sigma^{\theta\mu} \eta^{(\nu\sigma} p^\rho p^{\lambda)} 
p_\theta ~+~ \frac{(d+4)}{p^2} \sigma^{\theta(\nu} p^\sigma p^\rho p^{\lambda)}
p^\mu p_\theta \nonumber \\
&& -~ \sigma^{\theta(\nu} \eta^{\sigma\rho} p^{\lambda)} p^\mu p_\theta ~-~ 
(d+4) \sigma^{\mu(\nu} p^\sigma p^\rho p^{\lambda)} ~+~ p^2 \sigma^{\mu(\nu} 
\eta^{\sigma\rho} p^{\lambda)} \nonumber \\  
T^{\mu\nu\sigma\rho\lambda}_{2}(p) &=& 
-~ \frac{(d+2)(d+4)}{p^2} \sigma^{\theta\mu} p^\nu p^\sigma p^\rho p^\lambda 
p_\theta ~-~ p^2 \sigma^{\theta\mu} \eta^{(\nu\sigma} \eta^{\rho\lambda)}
p_\theta ~+~ (d+2) \sigma^{\theta\mu} \eta^{(\nu\sigma} p^\rho p^{\lambda)} 
p_\theta \nonumber \\ 
&& -~ \frac{(d+4)}{p^2} \sigma^{\theta(\nu} p^\sigma p^\rho p^{\lambda)}
p^\mu p_\theta ~+~ \sigma^{\theta(\nu} \eta^{\sigma\rho} p^{\lambda)} p^\mu 
p_\theta \nonumber \\  
T^{\mu\nu\sigma\rho\lambda}_{3}(p) &=& 
-~ \frac{4(d+4)}{p^2} \sigma^{\theta\mu} p^\nu p^\sigma p^\rho p^\lambda 
p_\theta ~+~ 2 \sigma^{\theta\mu} \eta^{(\nu\sigma} p^\rho p^{\lambda)} 
p_\theta ~-~ p^2 \sigma^{\theta(\nu} \eta^{|\mu|\sigma} \eta^{\rho\lambda)} 
p_\theta \nonumber \\
&& -~ \frac{(d+4)(d+5)}{p^2} \sigma^{\theta(\nu} p^\sigma p^\rho p^{\lambda)} 
p^\mu p_\theta ~+~ (d+4) \sigma^{\theta(\nu} \eta^{|\mu|\sigma} p^\rho 
p^{\lambda)} p_\theta ~.  
\end{eqnarray} 
Finally, it is worth observing that in the two loop computation of the 
anomalous dimensions of \cite{5,6,7,27,28,29,30,31} the Feynman rules were 
constructed by introducing a null vector $\Delta_\mu$ which projected out a 
part of the Green's function which had a non-zero tree term and therefore 
allowed for the extraction of the anomalous dimension. Since the lattice 
matching specifically requires information on the full Lorentz structure of the
Green's function we cannot follow that approach as the null vector would 
exclude access to several of the amplitudes which are extracted from the data 
by choosing different momentum configurations. 

\sect{Operators.}
In this section we list the full form of the four operators we considered
which satisfy the symmetrization and tracelessness properties. For the Wilson
operators we have 
\begin{equation}
{\cal O}_W^{\mu\nu\sigma} ~=~ W^{\mu\nu\sigma} ~-~ \frac{1}{(d+2)}
\eta^{(\mu\nu} W^{\sigma)\theta}_{~~~~\theta} 
\end{equation}  
for $n$~$=$~$3$ where 
\begin{equation}
W^{\mu\nu\sigma} ~=~ \frac{1}{6} \bar{\psi} \gamma^{(\mu} D^\nu D^{\sigma)} 
\psi
\end{equation} 
and for $n$~$=$~$4$ 
\begin{equation}
{\cal O}_W^{\mu\nu\sigma\rho} ~=~ W^{\mu\nu\sigma\rho} ~-~ \frac{1}{(d+4)}
\eta^{(\mu\nu} W^{\sigma\rho)\theta}_{~~~~~\theta} ~+~ \frac{1}{(d+2)(d+4)}
\eta^{(\mu\nu} \eta^{\sigma\rho)} W^{\theta~\phi}_{~\,\theta~\phi} 
\end{equation}  
where
\begin{equation}
W^{\mu\nu\sigma\rho} ~=~ \frac{1}{24} \bar{\psi} \gamma^{(\mu} D^\nu D^\rho 
D^{\sigma)} \psi ~. 
\end{equation} 
For the transversity operators, which have one fewer traceless conditions
compared to the Wilson operators, the full operators are more involved. For
$n$~$=$~$3$ we have 
\begin{eqnarray}
{\cal O}_T^{\mu\nu\sigma\rho} &=& T^{\mu\nu\sigma\rho} ~-~ 
\frac{(d^2+4d+2)}{d(d+2)(d+4)} T^{\mu(\nu|\theta}_{~~~~~~\,\theta|} 
\eta^{\sigma\rho)} ~-~ \frac{2}{d(d+2)(d+4)} T^{(\nu|\mu\theta}_{~~~~~~\theta|}
\eta^{\sigma\rho)} \nonumber \\ 
&& +~ \frac{1}{d(d+4)} T^{(\nu\sigma|\theta}_{~~~~~~\theta} 
\eta^{\mu|\rho)} ~+~ \frac{2}{d(d+4)} T^{\theta\mu(\nu}_{~~~~~|\theta|} 
\eta^{\sigma\rho)} \nonumber \\
&& -~ \frac{(d+2)}{d(d+4)} T^{\theta(\nu\sigma}_{~~~~~|\theta} 
\eta^{\mu|\rho)} ~+~ \frac{1}{(d+2)(d+4)} T^{\theta\phi}_{~~~\theta\phi} 
\eta^{\mu(\nu} \eta^{\sigma\rho)}  
\end{eqnarray}
where 
\begin{equation}
T^{\mu\nu\sigma\rho} ~=~ \frac{1}{6} \bar{\psi} \sigma^{\mu(\nu} D^\sigma 
D^{\rho)} \psi ~.
\end{equation} 
Finally, for $n$~$=$~$4$ the fully symmetrized and traceless operator is 
\begin{eqnarray}
{\cal O}_T^{\mu\nu\sigma\rho\lambda} &=& T^{\mu\nu\sigma\rho\lambda} ~-~ 
\frac{(d^2+7d+8)}{(d+1)(d+4)(d+6)} T^{\mu(\nu\sigma|\theta}_{~~~~~~~\theta|} 
\eta^{\rho\lambda)} \nonumber \\
&& +~ \frac{(d+5)}{(d+1)(d+4)(d+6)} T^{\mu\theta~\,\phi}_{~~~\theta~\,\phi} 
\eta^{(\nu\sigma} \eta^{\rho\lambda)} ~-~ \frac{2}{(d+1)(d+4)(d+6)} 
T^{(\nu|\mu|\sigma|\theta}_{~~~~~~~~~\theta|} \eta^{\rho\lambda)} \nonumber \\
&& +~ \frac{1}{(d+1)(d+6)} T^{(\nu\sigma\rho|\theta}_{~~~~~~~\theta} 
\eta^{\mu|\lambda)} ~-~ \frac{1}{(d+1)(d+4)(d+6)} 
T^{(\nu|\theta~\,\phi}_{~~~~\,\theta~\,\phi} \eta^{\mu|\sigma} 
\eta^{\rho\lambda)} \nonumber \\
&& +~ \frac{2}{(d+1)(d+6)} T^{\theta\mu(\nu\sigma}_{~~~~~~\,|\theta|} 
\eta^{\rho\lambda)} ~-~ \frac{4}{(d+1)(d+4)(d+6)} 
T^{\theta\mu~\,\phi}_{~~~\theta~~\phi} \eta^{(\nu\sigma} \eta^{\rho\lambda)}
\nonumber \\
&& -~ \frac{(d+4)}{(d+1)(d+6)} T^{\theta(\nu\sigma\rho}_{~~~~~~|\theta} 
\eta^{\mu|\lambda)} ~+~ \frac{(d+2)}{(d+1)(d+4)(d+6)} 
T^{\theta(\nu|~~\phi}_{~~~~~\theta~~\phi} \eta^{\mu|\sigma} \eta^{\rho\lambda)} 
\end{eqnarray}
where 
\begin{equation}
T^{\mu\nu\sigma\rho\lambda} ~=~ \frac{1}{24} \bar{\psi} \sigma^{\mu(\nu} D^\rho
D^{\sigma} D^{\lambda)} \psi ~. 
\end{equation}

\end{document}